\documentclass[aps,prl,final,twocolumn,letterpaper, superscriptaddress, showkeys]{revtex4-2}

\makeatletter
\AtBeginDocument{\let\LS@rot\@undefined}
\makeatother

\def\supplementfilename{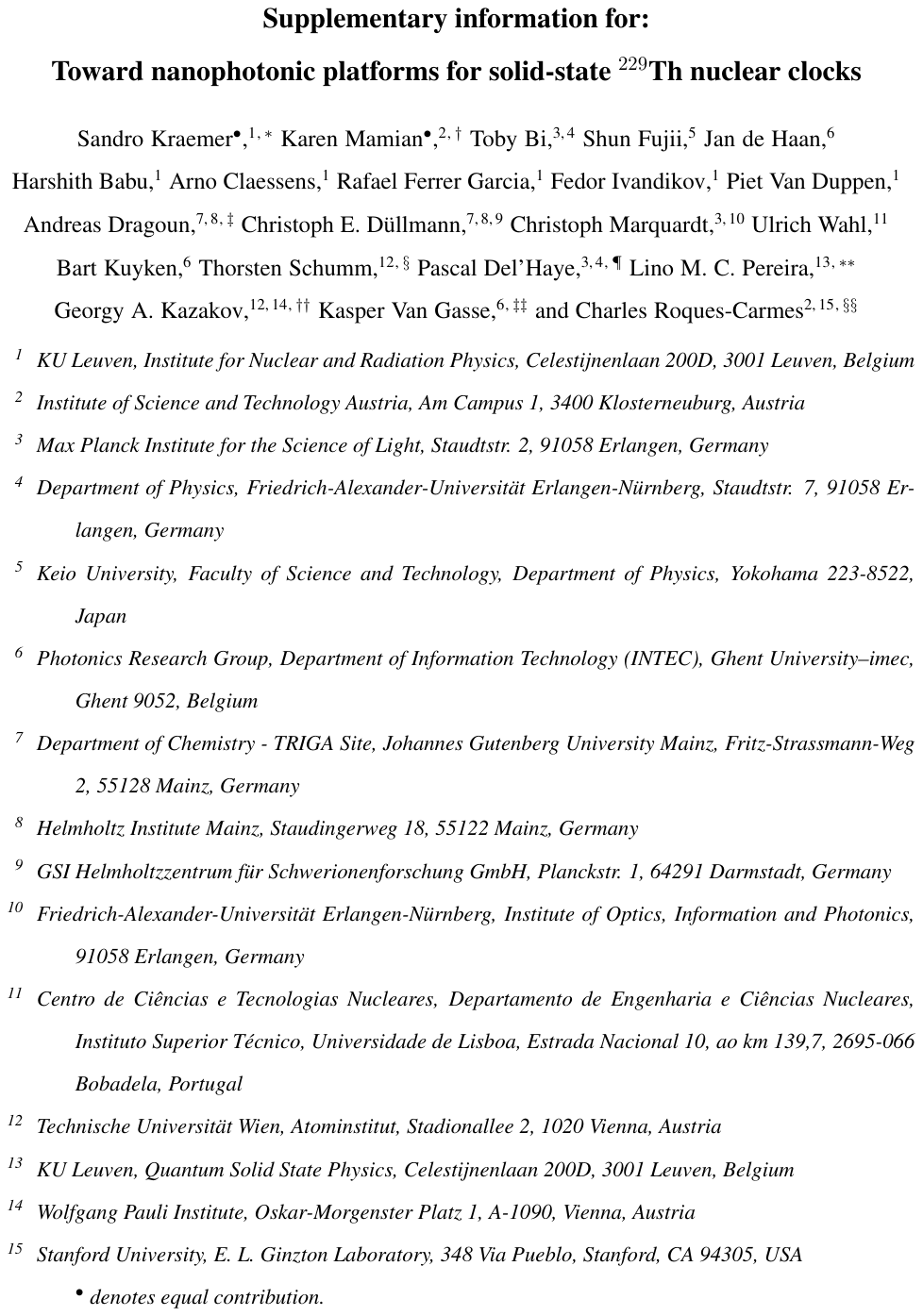}

\pdfximage{\supplementfilename}
\def\numbersupplementpages{\the\pdflastximagepages}

\newif\ifarXiv
\arXivtrue

\usepackage{pdfpages}
\usepackage{pgffor}
\usepackage{pgf}
\usepackage{graphicx}   
\usepackage{import} 
\usepackage{placeins}
\usepackage{epstopdf}
\usepackage{amsmath} 
\usepackage{float} 
\usepackage{bm}
\usepackage{amssymb}
\usepackage{quotes}
\usepackage{indentfirst}
\usepackage{color}
\usepackage[utf8]{inputenc}
\usepackage{transparent}
\usepackage{dcolumn}
\usepackage{braket}
\usepackage{multirow}
\usepackage{cancel} 
\usepackage{mdframed}
\usepackage{color}
\usepackage{dsfont}

\usepackage{slashed}
\usepackage{siunitx}

\usepackage{soul, color} 
\soulregister\ref{7}  
\soulregister\cite{7} 
\renewcommand{\st}[1]{}

\usepackage{xr}

\makeatletter
\newcommand*{\addFileDependency}[1]{
  \typeout{(#1)}
  \@addtofilelist{#1}
  \IfFileExists{#1}{}{\typeout{No file #1.}}
}
\makeatother

\newcommand*{\myexternaldocument}[2][]{%
    \externaldocument[#1]{#2}%
    \addFileDependency{#2.aux}%
}

\myexternaldocument[SI-]{si}
\usepackage{textcomp} 
\usepackage{xifthen}
\usepackage{xcolor}
\usepackage{etoolbox}
\newboolean{togglechanges} 

\setboolean{togglechanges}{false}

\newcommand{\comment}[1]{\ifbool{togglechanges}
    {#1}  
    {\textcolor{blue}{#1}}}
\def\Th{^{229}\text{Th}}

\usepackage{bibentry}

\usepackage{textcomp} 

\renewcommand*\rmdefault{ptm}

\makeatletter
\renewcommand{\fnum@figure}{\textbf{Fig.~\thefigure}}
\makeatother

\usepackage{dcolumn}
\usepackage{comment}

\usepackage{xcolor}
\usepackage{hyperref}

\renewcommand{\figurename}{Figure}
\renewcommand{\tablename}{Table}

\begin{document}

\title{Toward nanophotonic platforms for solid-state \texorpdfstring{${}^{229}$Th}{229Th} nuclear clocks}

\author{Sandro~Kraemer$^{\bullet}$}
\email{sandro.kraemer@kuleuven.be}
\thanks{These authors contributed equally.}
\affiliation{KU Leuven, Institute for Nuclear and Radiation Physics, Celestijnenlaan 200D, 3001 Leuven, Belgium}

\author{Karen~Mamian$^{\bullet}$}
\email{karen.mamian@ist.ac.at}
\thanks{These authors contributed equally.}
\affiliation{Institute of Science and Technology Austria, Am Campus 1, 3400 Klosterneuburg, Austria}

\author{Toby~Bi}
\affiliation{Max Planck Institute for the Science of Light, Staudtstr. 2, 91058 Erlangen, Germany}
\affiliation{Department of Physics, Friedrich-Alexander-Universität Erlangen-Nürnberg, Staudtstr. 7, 91058 Erlangen, Germany}

\author{Shun~Fujii}
\affiliation{Keio University, Faculty of Science and Technology, Department of Physics, Yokohama 223-8522, Japan}

\author{Jan~de~Haan}
\affiliation{Photonics Research Group, Department of Information Technology (INTEC), Ghent University–imec, Ghent 9052, Belgium}

\author{Harshith~Babu}
\affiliation{KU Leuven, Institute for Nuclear and Radiation Physics, Celestijnenlaan 200D, 3001 Leuven, Belgium}

\author{Arno~Claessens}
\affiliation{KU Leuven, Institute for Nuclear and Radiation Physics, Celestijnenlaan 200D, 3001 Leuven, Belgium}

\author{Rafael~Ferrer~Garcia}
\affiliation{KU Leuven, Institute for Nuclear and Radiation Physics, Celestijnenlaan 200D, 3001 Leuven, Belgium}

\author{Fedor~Ivandikov}
\affiliation{KU Leuven, Institute for Nuclear and Radiation Physics, Celestijnenlaan 200D, 3001 Leuven, Belgium}

\author{Piet~Van~Duppen}
\affiliation{KU Leuven, Institute for Nuclear and Radiation Physics, Celestijnenlaan 200D, 3001 Leuven, Belgium}

\author{Andreas~Dragoun}
\altaffiliation{Present address: Forschungszentrum Jülich, Jülich, Germany}
\affiliation{Department of Chemistry - TRIGA Site, Johannes Gutenberg University Mainz, Fritz-Strassmann-Weg 2, 55128 Mainz, Germany}
\affiliation{Helmholtz Institute Mainz, Staudingerweg 18, 55122 Mainz, Germany}

\author{Christoph~E.~D\"{u}llmann}
\affiliation{Department of Chemistry - TRIGA Site, Johannes Gutenberg University Mainz, Fritz-Strassmann-Weg 2, 55128 Mainz, Germany}
\affiliation{Helmholtz Institute Mainz, Staudingerweg 18, 55122 Mainz, Germany}
\affiliation{GSI Helmholtzzentrum für Schwerionenforschung GmbH, Planckstr. 1, 64291 Darmstadt, Germany}

\author{Christoph~Marquardt}
\affiliation{Max Planck Institute for the Science of Light, Staudtstr. 2, 91058 Erlangen, Germany}
\affiliation{Friedrich-Alexander-Universität Erlangen-Nürnberg, Institute of Optics, Information and Photonics, 91058 Erlangen, Germany}

\author{Ulrich~Wahl}
\affiliation{Centro de Ciências e Tecnologias Nucleares, Departamento de Engenharia e Ciências Nucleares, Instituto Superior Técnico, Universidade de Lisboa, Estrada Nacional 10, ao km 139,7, 2695-066 Bobadela, Portugal}

\author{Bart~Kuyken}
\affiliation{Photonics Research Group, Department of Information Technology (INTEC), Ghent University–imec, Ghent 9052, Belgium}

\author{Thorsten~Schumm}
\email{thorsten.schumm@tuwien.ac.at}
\affiliation{Technische Universität Wien, Atominstitut, Stadionallee 2, 1020 Vienna, Austria}

\author{Pascal~Del'Haye}
\email{pascal.delhaye@mpl.mpg.de}
\affiliation{Max Planck Institute for the Science of Light, Staudtstr. 2, 91058 Erlangen, Germany}
\affiliation{Department of Physics, Friedrich-Alexander-Universität Erlangen-Nürnberg, Staudtstr. 7, 91058 Erlangen, Germany}

\author{Lino~M.~C.~Pereira}
\email{lino.pereira@kuleuven.be}
\affiliation{KU Leuven, Quantum Solid State Physics, Celestijnenlaan 200D, 3001 Leuven, Belgium}

\author{Georgy~A.~Kazakov}
\email{kazakov.george@gmail.com}
\affiliation{Technische Universität Wien, Atominstitut, Stadionallee 2, 1020 Vienna, Austria}
\affiliation{Wolfgang Pauli Institute, Oskar-Morgenster Platz 1, A-1090, Vienna, Austria}

\author{Kasper~Van~Gasse}
\email{kasper.vangasse@ugent.be}
\affiliation{Photonics Research Group, Department of Information Technology (INTEC), Ghent University–imec, Ghent 9052, Belgium}

\author{Charles~Roques-Carmes}
\email{crc@ista.ac.at}
\affiliation{Institute of Science and Technology Austria, Am Campus 1, 3400 Klosterneuburg, Austria}
\affiliation{Stanford University, E. L. Ginzton Laboratory, 348 Via Pueblo, Stanford, CA 94305, USA  \\ 
 $^{\bullet}$ denotes equal contribution.}

\keywords{nuclear clock, solid-state frequency reference, nanophotonics, quantum optics}

\maketitle

\textbf{While the $^{229}$Th nuclear isomer has recently been observed and laser-excited, converting optical nuclear manipulation into a chip-scale solid-state frequency standard remains an open challenge.
Here, we present a nanophotonic platform to realize an all-solid-state nuclear clock based on the low-energy isomeric transition of 
$^{229}$Th embedded in high-$Q$ fluoride photonic resonators. By coupling ensembles of thorium nuclei to confined optical modes, we show that resonant field build-up in the cavity can substantially enhance the nuclear excitation rate, enabling optical interrogation at practical laser intensities. We model the nuclei-photon interaction dynamics and outline a technological roadmap toward addressing this challenge, including resonator fabrication in fluoride crystals, thorium implantation, nuclear excitation with integrated lasers, and on-chip detection of vacuum-ultraviolet photons. As an initial proof of concept, we implant a crystalline fluoride whispering-gallery-mode resonator with $^{229}$Th and assess the impact of implantation-induced damage on resonator performance. Our platform leverages recent advances in materials integration and nanophotonics to chart a realistic route toward compact and scalable nuclear frequency standards.}

Atomic clocks provide humanity’s most precise measure of time and frequency, underpinning global navigation, communication, and quantum metrology~\cite{ludlow2015optical}. Optical atomic clocks based on electronic transitions in the visible and infrared ranges have now reached fractional uncertainties below $10^{-18}$~\cite{brewer_27mathrm_2019,aeppli_clock_2024}, 
but their size, complexity, and power requirements largely confine them to specialized laboratories. Meanwhile, integrated and chip-scale atomic clocks based on micro-electromechanical system (MEMS) technology achieve stabilities around $10^{-11}$~\cite{kitching2018chip}, still relying on gas-phase atomic ensembles in macroscopic cells. A solid-state frequency reference could overcome these constraints, enabling compact and robust time standards deployable far beyond the laboratory.

An emerging route toward ultrastable frequency standards is the nuclear clock based on the narrow nuclear transition of $^{229}$Th with an exceptionally low energy of $E_\mathrm{iso} = 8.4$~eV. Owing to the weak coupling of the nucleus to its environment, this transition is predicted to enable accuracies approaching $10^{-19}$~\cite{campbell_single-ion_2012}. The existence of a low-energy nuclear isomer in $^{229}$Th was first inferred from $\gamma$-spectroscopy in the 1970s \cite{kroger_features_1976}. Indirect measurements deduced excitation energies of $3.5(1)$~eV~\cite{helmer_excited_1994}, and later of $7.8(0.5)$~eV, corresponding to a wavelength near $160$~nm~\cite{beck_energy_2007,beck_improved_2009}. The concept of a nuclear clock based on this transition was introduced by Peik and Tamm in 2003~\cite{peik_nuclear_2003}.

Direct experimental access to the isomer was achieved in 2016 through the detection of conversion electrons following $^{233}$U $\alpha$-decay populating the metastable state $^{229m}$Th~\cite{von_der_wense_direct_2016}. Subsequent experiments determined its lifetime, energy, and electromagnetic moments~\cite{seiferle_lifetime_2017,thielking_laser_2018,seiferle_energy_2019}. More recently, implantation of radioactive ion beams into wide-bandgap crystals at CERN-ISOLDE enabled the observation of the radiative decay and demonstrated the feasibility of the solid-state approach~\cite{verlinde_alternative_2019,kraemer_observation_2023}. The improved wavelength value of $148.71(42)$~nm lead in 2024 to the first laser excitation of the nuclear transition ~\cite{tiedau_laser_2024}, measured at $148.3821(5)$~
nm. Vacuum-ultraviolet frequency-comb spectroscopy subsequently linked the nuclear transition to the $^{87}$Sr optical clock transition while resolving the temperature-dependent quadrupole splitting~\cite{zhang_frequency_2024,higgins_temperature_2025,ooi_frequency_2026}. These advances establish the foundation for a solid-state nuclear clock, in which thorium nuclei are embedded in wide-bandgap crystals transparent at $148.4$~nm, suppressing internal conversion and enabling scalable architectures~\cite{rellergert_constraining_2010,kazakov_performance_2012}. Radiative decay has now been observed in several host materials including CaF$_2$, MgF$_2$, LiSrAlF$_6$, and ThF$_4$~\cite{elwell_laser_2024,zhang_229thf4_2024,pineda_radiative_2025}.

A promising route to make the nuclear clock compact and scalable lies in nanophotonics, which enables control of light-matter interactions at the nanoscale~\cite{joannopoulos2011photonic}. High-$Q$ resonators with small mode volumes generate large intracavity fields with relatively low input powers and modify the local density of optical states, enhancing emission rates via the Purcell effect while enabling efficient in- and out-coupling and chip-scale routing. In parallel, low-loss integrated photonic platforms now support ultraviolet frequency conversion, compact microcomb sources, and scalable passive and active photonic elements for precision metrology~\cite{Liu2021,Wu2025}.

Here, we propose and validate a path towards an all-solid-state nuclear clock based on embedding $^{229}$Th nuclei into high-$Q$ whispering-gallery-mode (WGM) nanophotonic resonators. We first develop a theoretical framework describing the interaction between an ensemble of thorium nuclei and confined optical modes in a photonic cavity, and use it to estimate the signal from the nuclear transition in different $\Th$ embedding scenarios. We then outline a technological roadmap toward realizing this concept, encompassing vacuum ultraviolet (VUV) photonic resonator design and fabrication, integration of thorium into suitable host crystals, development of narrow-linewidth ultraviolet and VUV laser sources for direct or two-photon excitation, as well as on-chip detection strategies for the emitted nuclear photons. As an initial proof-of-concept, we successfully implant a magnesium fluoride photonic resonator with $^{229}$Th to establish the compatibility of thorium implantation with crystalline fluoride resonators. 
Taken together, these results establish the technological feasibility of a scalable, robust, and fully integrated solid-state frequency reference that harnesses the advantages of a nuclear transition.

\section{Physics of nuclei-photon interactions in a nanophotonic resonator}

\begin{figure}
    \includegraphics[width=1\linewidth]{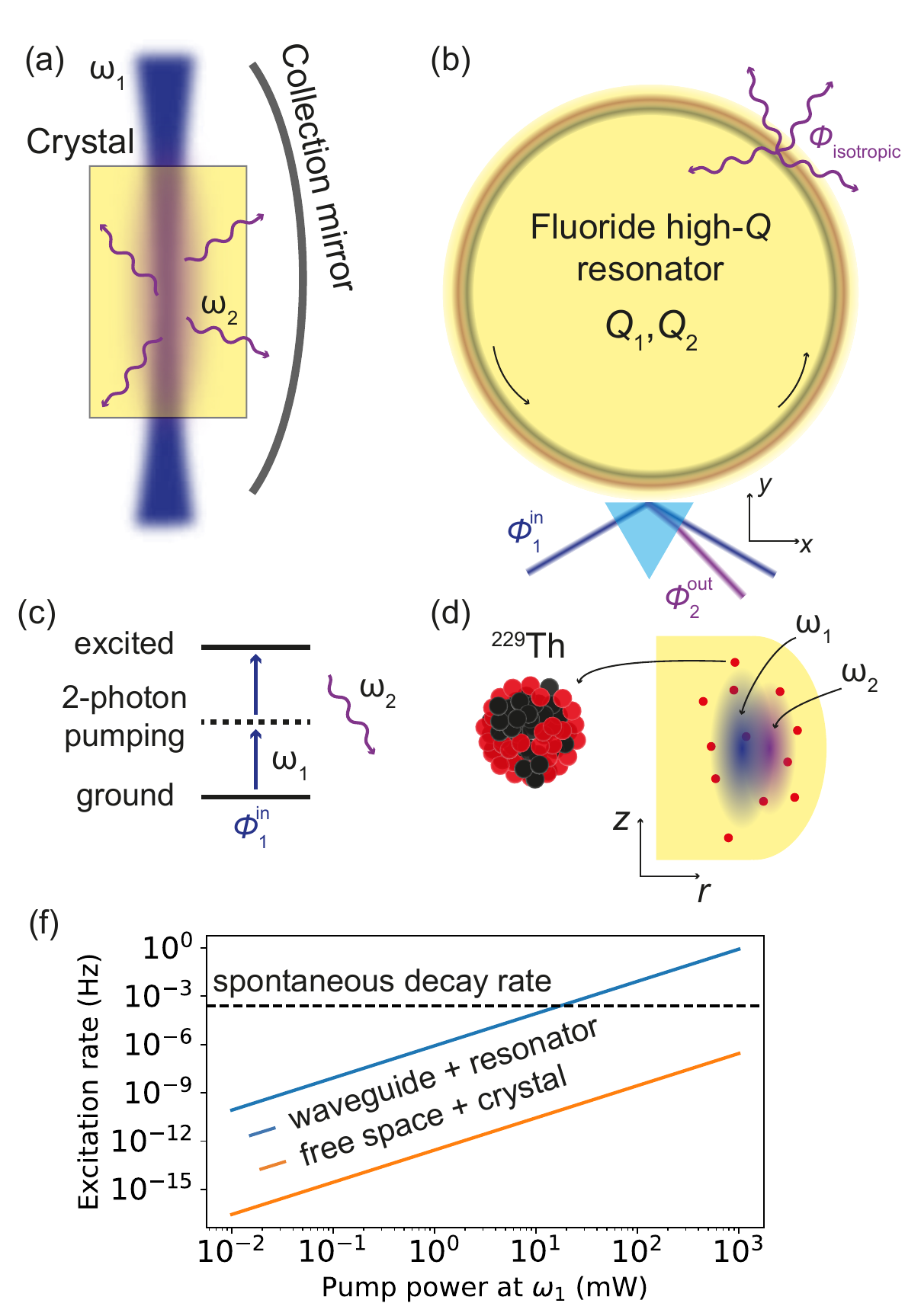}
    \caption{\textbf{Concept of an all solid-state nuclear clock.} (a) Conventional pumping scheme for a thorium-doped crystal as used in \cite{tiedau_laser_2024,elwell_laser_2024,zhang_frequency_2024}. (b) Proposed high-$Q$ thorium-doped fluoride optical resonator. (c) We use a two-photon pumping scheme leveraging resonances with corresponding $Q$-factors $Q_1$ at $\omega_2=E_\text{iso} / \hbar$ and  $Q_2$ at $\omega_1 = \omega_2/2$. (d) Overlap of the thorium nuclei with the optical modes. (f) Nuclear two-photon excitation rates as a function of laser power for the free space (a) and integrated (b) approaches.}
    \label{fig:concept}
\end{figure}

We first outline the key physical ingredients that underpin our approach: the interaction between embedded \(^{229}\)Th nuclei and confined optical modes in high-\(Q\) photonic resonators made of wide-bandgap UV- and VUV-transparent materials, and the mechanisms enabling coherent optical pumping of the nuclear transition (shown in Fig.~\ref{fig:concept} (a-d)).  

\subsection{\label{sec:concept}Basics of high-\texorpdfstring{$Q$ w} whispering-gallery-mode (WGM) resonators}

Whispering-gallery-mode (WGM) resonators confine light through continuous total internal reflection along a curved dielectric interface with minimal losses, allowing optical fields to circulate many times within a compact volume. The resulting optical confinement and long photon lifetimes yield high intracavity intensities and narrow linewidths, providing an ideal environment for enhancing light-matter interactions. Such resonators can be realized in various geometries -- including microspheres, disks, and toroids -- and support modes characterized by an azimuthal propagation constant and a discrete free spectral range (FSR) set by the resonator circumference and refractive index~\cite{vahala2003optical, matsko2006optical, ilchenko2006optical}. 

The quality factor ($Q$-factor) of the resonator is $Q = \omega/\kappa_{i}$ where $\kappa_{i}$ is the total photon decay rate. The intrinsic loss rate includes contributions from absorption in the bulk and at the surface, scattering due to surface roughness, and radiation due to weak confinement. In a millimeter-scale WGM resonator, however, the intrinsic losses are expected to be dominated by surface scattering, scaling as $Q \propto \lambda^{3} \sigma^{-2}$ with wavelength $\lambda$ and surface roughness $\sigma$ \cite{gorodetsky_rayleigh_2000}. Because crystalline materials cannot be reflowed without degrading their lattice quality, ultrahigh-$Q$ single-crystal resonators are instead fabricated directly from bulk crystals using single-point diamond turning~\cite{grudinin2008single}, femtosecond-laser ablation~\cite{Lin2015}, or precision grinding and polishing on a lathe~\cite{grudinin2006ultra}, followed by fine polishing steps to minimize surface roughness and suppress scattering losses~\cite{fujii_all-precision-machining_2020,yang2025systemaic}. To enable operation in the UV and VUV regimes relevant to the $^{229}$Th nuclear spectroscopy, single-crystal fluoride materials are preferred, as they provide broad transparency windows into the VUV~\cite{dessovic_229thorium-doped_2014,pimon_dft_2020,morgan_229th-doped_2024}, low scattering losses, and high optical damage thresholds. Indeed, WGM resonators based on MgF$_2$ and CaF$_2$ have demonstrated ultrahigh quality factors exceeding \(10^8\) in the visible~\cite{grudininfundamental2007, savchenkovselfinjection2019}, up to \(10^{11}\) in the near-infrared~\cite{Savchenkov2007}, and above \(10^8\) in the mid-infrared~\cite{savchenkov2015generation}, enabling a wide range of photonic applications. The strong optical confinement and long photon lifetimes in these resonators support enhanced nonlinear interactions, including frequency upconversion~\cite{furst_naturally_2010,strekalov_nonlinear_2016} and harmonic generation~\cite{breunig_three_2016}.

Light can be coupled efficiently into WGM resonators through several evanescent coupling schemes, including tapered optical fibers~\cite{spillane_ideality_2003}, prisms~\cite{gorodetsky_optical_1999}, angle-polished fibers~\cite{lin2017nonlinear}, or integrated waveguides~\cite{Anderson2018,liulowloss2018}. By controlling the coupling gap, the external coupling rate can be tuned from under- to over-coupled regimes, with critical coupling achieved when the external and intrinsic losses are balanced. Multiple out-coupling channels can be introduced by using multiple couplers (i.e., tapered fibers or prisms) with coupling ratios optimized for different wavelengths. This flexibility allows efficient excitation and extraction of optical fields across multiple spectral bands, including the UV and VUV, and provides a robust interface between free-space laser systems and integrated photonic devices.

The combination of high $Q$-factors, relatively small mode volumes, and broad spectral coverage makes crystalline fluoride WGM resonators uniquely suited for coupling to embedded nuclear transitions. Their ability to concentrate optical energy in sub-millimeter volumes while maintaining low absorption opens a realistic pathway to enhance the interaction strength between confined VUV photons and \(^{229}\)Th nuclei in an all–solid-state nuclear clock architecture. For readers outside the field, we review the basic concepts of nanophotonics and coupled-mode theory in the Supplementary Information (SI), Section~\ref*{SI-si:nanoCMT}.

\subsection{Resonant laser excitation of the nuclear transition}

Interaction of monochromatic laser light with photon energy \(\hbar \omega\) and intensity \(I\) can drive the transition between the ground and isomeric nuclear states. In the low-intensity limit, the excitation rate $R_\mathrm{exc}$ can be expressed via an effective cross section \(\sigma_i\) as
\begin{equation}
    R_\mathrm{exc} = \sigma_i \frac{I}{\hbar \omega},
\end{equation}
The dominant coupling of $\Th$ nuclei to the electromagnetic field arises from the magnetic-dipole (M1) transition, with estimated probabilities in the range \(1.4\times10^{-4}~\mathrm{s^{-1}} \leq \Gamma_\mathrm{M1} \leq 6.5\times10^{-4}~\mathrm{s^{-1}}\)~\cite{minkov_reduced_2017,tkalya_radiative_2015}. Hereinafter we adopt \(\Gamma_\mathrm{M1} = 4.0\times10^{-4}~\mathrm{s^{-1}}\), corresponding to a half-life of $1740$~s in vacuum. The weaker electric-quadrupole (E2) component contributes with \(\Gamma_\mathrm{E2} \approx 1.1\times10^{-13}~\mathrm{s^{-1}}\)~\cite{minkov_reduced_2017,ruchowska_nuclear_2006}. 

\textbf{One-photon excitation.} For light resonant with the nuclear transition ($\hbar \omega=E_{eg}=E_{\rm iso}$, where $e$ and $g$ denote specific sublevels of the isomer and ground states in the local crystal lattice environment), the absorption cross section of the M1 channel can be approximated as~\cite{von_der_wense_theory_2020}:
\begin{equation}
\begin{split}
    \sigma_\mathrm{M1-1ph} =& \frac{2\pi c^2 \hbar^2}{E_\mathrm{iso}^2} \frac{\Gamma^\mathrm{M1}_{eg}}{\Gamma_{L}} \\
    \approx& \frac{1.4\times10^{-18}~\mathrm{m^2\,s^{-1}}}{\Gamma_{L}} |C_{I_gm_g1q}^{I_em_e}|^2,
    \label{eq:2}
\end{split}
\end{equation}
where $m_g$ and $m_e$ are magnetic quantum numbers of the ground and the excited states, $C_{I_gm_g1q}^{I_em_e}$ is the Clebsch-Gordan coefficient (where $\Gamma^{\rm M1}_{eg}=\Gamma_{\rm M1}|C_{I_gm_g1q}^{I_em_e}|^2$ is the partial decay rate), and $q=m_e-m_g$. Hereafter, we consider the  $m_g=\pm1/2 \rightarrow m_e=\pm1/2$ $\pi$-transitions with $|C_{I_gm_g1q}^{I_em_e}|^2=2/5$. 
Hence, the cross section scales inversely with the laser linewidth \(\Gamma_{L}\) and is independent of intensity $I$.

\textbf{Two-photon excitation.} Several schemes relying on multi-photon excitations have been proposed to overcome the limitations in laser technology at the $148.4$~nm wavelength (Fig.~\ref{fig:concept}~(c)). We first consider direct single-color two-photon absorption of the nucleus (such that $2 \hbar \omega = E_\text{iso}$, and we restrict our consideration only to a pair of sublevels $g$ and $e$ of the ground and the isomeric state). The resulting cross-section reads as
\begin{multline}
    \sigma_\mathrm{M1-2ph} = \frac{\pi \hbar^4 e^2}{2 \epsilon_0 c m_p^2} \frac{I}{E_\mathrm{iso}^4} \frac{\Gamma^\mathrm{M1}_{eg}}{2\Gamma_{L}} \left(g_e m_e - g_g m_g\right)^2 \\
    \approx 3.0 \times 10^{-50} \text{ }\mathrm{m}^4 \mathrm{s}^{-1}\mathrm{W}^{-1} \frac{I}{2\Gamma_{L}}|C_{I_gm_g1q}^{I_em_e}|^2,
\end{multline}
where $m_p$ is the mass of the proton, $g_e=-0.252$ ($g_g=0.146$) and $m_e=3/2$ ($m_g=5/2$) are the isomeric (ground) state \textit{g}-factor and magnetic quantum numbers respectively~\cite{von_der_wense_theory_2020,Romanenko12}. It should be noted that the cross sections involving two-photon processes depend on the intensity $I$ of the laser light. With $ 10^{10} \mathrm{\text{ }Wm}^{-2}$ of laser intensity and $\Gamma_{L} = 2 \pi \times 1$~kHz, the direct two-photon absorption cross $\sigma_\mathrm{M1-2ph}$ section of the nucleus is more than 22 orders of magnitude smaller than the direct one-photon absorption cross section $\sigma_\mathrm{M1-1ph}$.
\begin{figure*}
    \centering  
\includegraphics[width=1.0\linewidth]{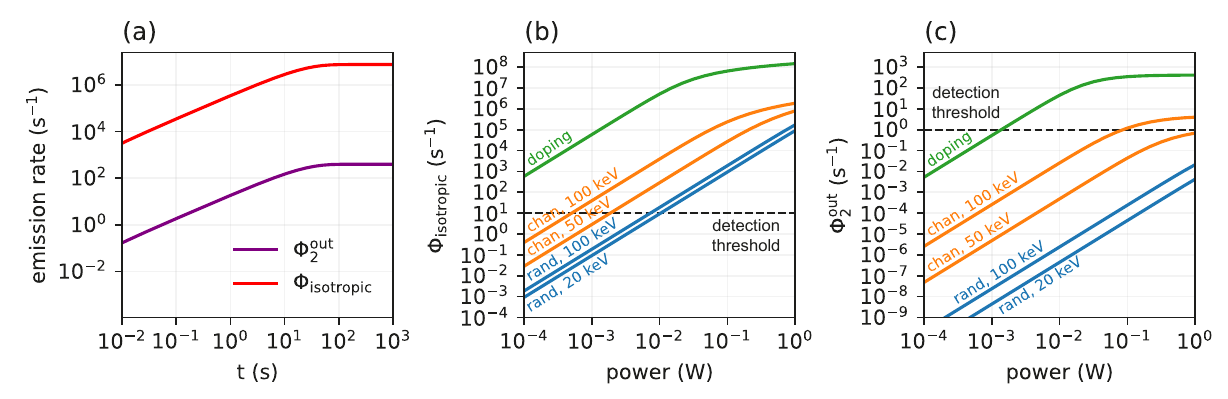}
    
    \caption{\textbf{Two-photon pumping of thorium nuclei embedded in a high-$Q$ photonic resonator.} (a) Calculated output photon flux dynamics of the $148$~nm mode ($\Phi^{\text{out}}_2$) and isotropic emission rate ($\Phi_{\text{isotropic}}$) with a pump power $P_1 = 100$~mW, for the case of maximum mode overlaps and uniform $\Th$ doping. (b, c) Steady-state dependence on pump power and implantation methods. Steady-state $\Phi_{\text{isotropic}}$ (b) and $\Phi^{\text{out}}_2$ (c) dependence on input power $P_1$ of the $296$~nm mode for randomly oriented implantation (blue) and implantation aligned with a major crystal axis, so-called channeled implantation, (orange) of $\Th$ (implantation fluence of $10^{13}$~cm${}^{-2}$), as well as doping of the full resonator (doping concentration of $2\times10^{17}$~cm${}^{-3}$) using simulated mode field distributions for the resonator shown in Fig.~\ref{fig:implantation}~(a). Black dashed lines represent the detection thresholds: for the cavity output case, a threshold of $1$~s${}^{-1}$ is assumed, while for the isotropic spontaneous nuclear emission we take a threshold of $10$~s${}^{-1}$, taking into account finite solid angle coverage.} 
    \label{fig:2photon}
\end{figure*}
A more promising route to realize two-photon driving of the nuclear transition relies on the coupling of the nucleus to the crystalline host. The laser beam polarizes the crystalline medium and couples to the nucleus via the electric field gradient and the nuclear quadrupole moment. This interaction is referred to as the optonuclear quadrupolar effect (ONQ) \cite{xu_solid-state_2023, xu2023two}. The resulting cross section $\sigma_\mathrm{ONQ}$ can be approximated as (e.g., in CaF$_2$):
\begin{multline}
    \sigma_\mathrm{ONQ} = \\ \frac{450 \hbar^4 c^3 e^6}{2 \pi^2  \epsilon_0^3 n^2a_0^2} \frac{1} {\left(E_\mathrm{bg}-E_\mathrm{iso}\right)^2 \left(E_\mathrm{bg}-\frac{E_\mathrm{iso}}{2}\right)^2} \frac{I}{E_\mathrm{iso}^4} \frac{\Gamma_\mathrm{E2}}{2\Gamma_{L}} 
    \\
    \approx 1.8 \times 10^{-38} \text{ } \mathrm{m}^4\mathrm{s}^{-1}\mathrm{W}^{-1} \frac{I}{2\Gamma_{L}} ,
\end{multline}
where $E_\mathrm{bg}$ is the electronic bandgap energy of the surrounding medium with $E_\mathrm{bg} > E_\text{iso}$ (see the derivation and discussion in the 
SI, \ref*{SI-si:onq_derivation}). Importantly, according to these calculations, this type of nucleus-photon interaction may be many orders of magnitude stronger than direct two-photon absorption. Additional two-color two-photon absorption schemes, relying on the coupling of electrons to the nucleus and commonly referred to as "electron-bridge excitation", have also been discussed in the literature \cite{Li2023}.

Notably, all of the above cross sections can be enhanced by embedding the nuclei into an engineered optical environment, such as a high-finesse optical resonator, where the spontaneous emission rate is increased by the corresponding Purcell factor \(F_P\). Whereas two-photon processes generally require high optical intensities, cavity-induced field confinement can reduce the required driving power by several orders of magnitude (Fig.~\ref{fig:concept}~(f)).

\vspace{0.8\baselineskip}

\subsection{Interactions between \texorpdfstring{$^{229}$Th n}  nuclei and photons in a photonic microresonator}

The interaction of nuclei embedded in a  high-$Q$ resonator with cavity photons may be modeled with a set of semi-classical rate equations:
\begin{widetext}
\begin{equation}
\begin{cases}
\displaystyle \frac{dN_1}{dt}&=-\kappa_1N_1+\Phi_{\rm in}^{\rm cav}+\displaystyle\sum\limits_{i=1}^{N_{C}}\frac{4g^2(\vec{r}_i)}{\Gamma_1}N_1^2S_{z,i}.
\\
\displaystyle\frac{dN_2}{dt}&=-\kappa_2N_2+\displaystyle\sum\limits_{i=1}^{N_C}\frac{4G^2(\vec{r}_i)}{\Gamma_2}N_2S_{z,i}+\displaystyle\sum\limits_{i=1}^{N_C}\frac{2G^2(\vec{r}_i)}{\Gamma_2}\bigl(\rho(\vec{r}_i) \Delta V_i+S_{z,i}\bigr),
\\
\displaystyle \frac{dS_{z,i}}{dt}&=-\left(\gamma+ \displaystyle \frac{4G^2(\vec{r}_i)}{\Gamma_2}\right)\bigl(S_{z,i}+\rho(\vec{r}_i) \Delta V_i \bigr)
    -\displaystyle \frac{4g^2(\vec{r}_i)}{\Gamma_1}N_1^2S_{z,i}
    -\frac{8G^2(\vec{r}_i)}{\Gamma_2}N_2S_{z,i}.
\end{cases}
\label{eq:rates}
\end{equation}
\end{widetext}
In the above equation, we divide the nuclear ensemble into $N_c$ clusters, where the nuclei are considered to be under identical conditions. $N_1, N_2$ are the cavity mode photon numbers at $\omega_1 (296 \text{ }\rm nm )= \omega_\mathrm{iso}/2$, $\omega_2 (148 \text{ }\rm nm) =\omega_\mathrm{iso}$, respectively, and $S_{z,i} = \sum_{j\in C_{i}}\langle \hat{\sigma}^j_{z}\rangle=N_{e,i}-N_{g,i}$ is the sum of individual nuclear population inversions $\langle \hat{\sigma}^j_{z}\rangle$ over all the nuclei in cluster $C_i$ with a volume of $\Delta V_i$; $N_{{\rm Th},i} = \rho(\vec{r}_i)\Delta V_i=N_{e,i} + N_{g,i}$ is the total number of nuclei in cluster $C_i$, $\rho(\vec{r})$ is the $\Th$ density distribution at the $m=\pm 1/2$ ground sublevel (the total density is three times larger due to the distribution over $m_g=-5/2, \ldots, 5/2$ sublevels); $\gamma = n_\text{CaF$_2$}^3 \times \Gamma_{\text{M1}}=1.59\times 10^{-3} \mathrm{\text{ }s^{-1}}$\cite{tiedau_laser_2024} is the spontaneous decay rate of $^{229}$Th nuclei through the dominant M1 channel in the case of a Th:CaF${}_2$ crystal. The emission rate into the cavity mode is given by the Purcell factor $\frac{4G^2(\vec{r})}{\Gamma_2}$, with $G(\vec{r})=\frac{\mu B_{zpf}(\vec{r})}{\hbar}$, where $B_{zpf}$ is the position-dependent vacuum magnetic field and $\mu = \sqrt{\frac{\sigma_\mathrm{M1-1ph} 3 \hbar c}{\mu_0 \omega_{\rm iso}}}$ is the magnetic dipole matrix element. We utilize the single-color two-photon pumping mechanism~\cite{xu_solid-state_2023} described in the previous section, which has gained attention thanks to the availability of laser sources around $300$~nm. It is characterized by an excitation rate $\frac{4g^2(\vec{r})}{\Gamma_1}$, where $g(\vec{r})=\frac{\langle g|D|m \rangle E_{zpf}^2(\vec{r})}{\hbar}$ with $E_{zpf}$ being the vacuum electric field and $\langle g|D|m \rangle = c \epsilon_0 n \sqrt{\frac{\sigma_{\rm ONQ} \Gamma_1}{2I}}$ the ONQ matrix element. 
Decoherence due to interaction with surrounding fluorine spins in the crystal is taken into account \cite{kazakov_performance_2012}. 

The mode at frequency $\omega_1$ is pumped at a photon rate $\Phi_{\text{in}}^\text{cav}$, and $\Phi_{2}^\text{out}$ photons per second exit via the waveguide port at frequency $\omega_2$. Additionally, photons within the cavity modes may decay via intrinsic losses $\kappa_i^\text{int}$, arising from effects such as material absorption or scattering, or through external coupling to an adjacent waveguide $\kappa_i^\text{ext}$: $\kappa_i = \kappa_i^\text{int} +\kappa_i^\text{ext} =\omega_i/Q_i $.  
The relevant basic concepts of nanophotonics, and a first-principle derivation of this set of equations can be found in the SI (\ref*{SI-si:nanoCMT}, \ref*{SI-si:derivations}).

The cavity output photon flux is $\Phi_{2}^\text{out}=\kappa_i^\text{ext} N_2 = \frac{1}{2}\kappa_i N_2$ (at critical coupling $\kappa_i=2\kappa_i^\text{ext}$, assumed in all simulations for both modes), and the isotropic emission rate $\Phi_{\text{isotropic}}=\gamma N_e$. To simulate the basic dynamics of output signals, we assume that the optical cavity modes overlap perfectly with each other and with the nuclear distribution (uniform $\Th$ doping case), thus neglecting all spatial distributions (constant $\rho, G,g$). 

Initially, the cavity is empty and all nuclei are in the ground state: $N_e (t=0)= 0$. Fig.~\ref{fig:2photon}~(a) shows the dynamics of the simulated cavity output photon flux and the isotropic emission rate at $148$~nm, with the quality factors $Q_1 =5 \times 10^6, Q_2=5\times10^6/8$ (implying the $Q \sim \lambda^{-3}$ scaling discussed above), and a uniform doping density of nuclei $\rho = 2\times 10^{17}$~cm$^{-3}$ (as is,~e.g., achievable at TU Wien \cite{tiedau_laser_2024}). A pump power of $100$~mW from a commercially available laser diode system results in an easily detectable output photon flux. All used parameters are presented in the Table \ref*{SI-table1}~(SI). We note that due to a relatively large field volume of these millimeter-scale resonators, the Purcell factor of the $148$~nm mode is much smaller than $1$. Therefore, it may be beneficial to collect isotropic spontaneous emission by a large aperture detector off-chip in initial realizations of this experiment. Additional dependencies of the cavity output flux on system parameters in this simple model are presented in Fig.~\ref*{SI-fig:cavity_sweeps}~(SI).

These calculations indicate that, with realistic parameters for cavity $Q$-factors, nuclear densities, and pump powers in the milliwatt range, the proposed resonator-assisted two-photon excitation scheme can yield measurable photon fluxes at the nuclear transition wavelength. The combination of cavity enhancement and optonuclear coupling provides order-of-magnitude improvements in excitation efficiency compared to direct free-space driving, underscoring the feasibility of an all-solid-state nuclear clock architecture.

\section{Fabrication and characterization of \texorpdfstring{$\Th$-I} implanted WGM resonators}

Realizing a nanophotonic nuclear clock requires a fabrication strategy that simultaneously maintains ultrahigh optical $Q$-factor and locates ${}^{229}$Th nuclei within the high-field region of the resonator mode. In this section, we establish ion implantation into single-crystal WGM resonators as a promising route to achieve that goal, supported by initial implantation experiments and modeling of the resulting signals as described below.

${}^{229}$Th can be introduced into the resonator either by using thorium-doped base material for the fabrication of the resonator (doping technique) or by ion implantation into a pre-fabricated resonator. 
Doping distributes the nuclei homogeneously over the bulk of the resonator and only a limited fraction overlaps with the region of the mode with high field strength. Doping concentrations on the order of $10^{18}$ cm${}^{-3}$ and narrow line nuclear spectroscopy with a $148$~nm laser on these crystals have been demonstrated~\cite{tiedau_laser_2024, zhang_229thf4_2024}. Fabrication requires, however, the processing of radioactive material during production. In a microresonator of $2$~mm diameter and $0.5$~mm height, the total $^{229}$Th activity at a doping density of $\rho=2\times 10^{17} \text { } \mathrm{cm}^{-3}$ will be below $1$~kBq, the exemption limit in many countries, enabling use of the integrated device without restrictions. Additionally, all thorium nuclei contribute to the comparably high activity of the resonator increasing the radioluminescence background while only a small fraction experiences a sufficient electromagnetic field intensity for two-photon excitation. The crystal growth parameters allow additionally to influence the microscopic environment at the ${}^{229}$Th dopant.

Introducing $^{229}$Th by ion implantation allows to separate the fabrication of high-$Q$ resonators made of high-purity crystals such as CaF$_2$, MgF$_2$ or BaF${}_2$ from the post-fabrication implantation of radioactive $^{229}$Th. Atomic nuclei are placed in a shallow layer close to the resonator surface with significant mode intensity. Achieving a strong overlap between the distribution of thorium nuclei and the electromagnetic mode is desirable, as the detection of nuclear decay photons grows compared to the radioluminescence background. 
\begin{figure}
    \centering
    \includegraphics[width=\columnwidth]{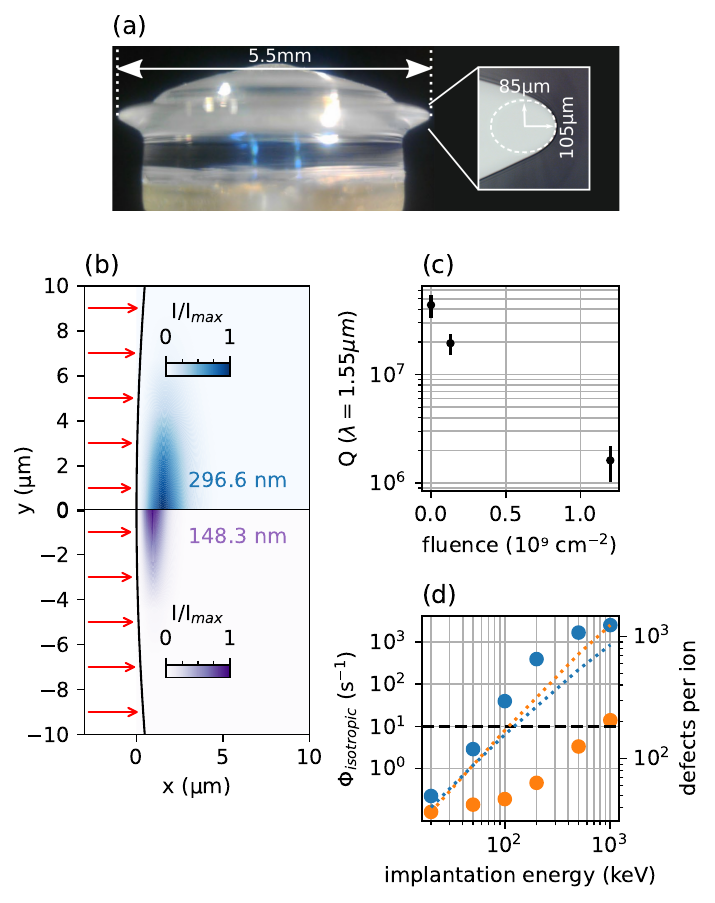}
        \caption{\textbf{Fabrication and characterization of $^{229}$Th-implanted WGM resonators.} The photograph in panel (a) shows a MgF$_2$ resonator implanted with ${}^{229}$Th. Simulated intensity distributions of the $296$~nm (upper panel) and $148$~nm (lower panel) modes in the resonator cross-section are shown in panel (b). The direction of implantation is shown by red arrows. (c) Experimental $Q$-factors measured at a wavelength of $1.55 \text{ }\mu$m for different implantation fluences of ${}^{229}$Th at 20 keV for the resonator. Panel (d) presents the  resulting projected isotropic nuclear fluorescence rate $\Phi_\mathrm{isotropic}$ assuming a coupled power of 1 mW, a $Q$-factor of $5\times10^6$ and a fluence of $1\times10^{13}$ cm${}^{-2}$ (dots) for randomly oriented (orange) and channeled (blue) implantation. The detection threshold is marked as dashed black line. Dashed orange and blue lines show defect creation per implanted atom for randomly oriented and channeled implantation, respectively.}
    \label{fig:implantation}
\end{figure}
As a first proof-of-principle step toward this approach, we measure the $Q$-factor of an (unimplanted) $z$-cut MgF${}_2$ resonator with a diameter of $ 5.5$~mm and a curvature radius of $105$~$\mu$m (Fig.~\ref{fig:implantation}~(a)) to be $4\times10^7$ at a wavelength of $\lambda=1.55\text{ }\mu$m.  The depth of the intensity maximum of the electromagnetic mode is influenced by the resonator shape, and for the fabricated resonator parameters it lies at $\approx 1.5~\mu$m and $\approx 0.9~\mu$m from the surface for $296$~nm and $148$~nm modes, respectively (see Fig. \ref{fig:implantation}~(a)). 
Typical implantation depth distributions for several implantation energies ranging from $20$~keV to $1$~MeV and corresponding simulated implantation depth distributions are shown in Fig.~\ref*{SI-fig:ImplantationDepth}~(SI) together with the electromagnetic field intensity in the horizontal plane. The higher implantation energies in this range place nuclei into regions with higher mode field intensity. Implantation into single-crystal resonators can be performed at a random incidence angle or parallel to one of the major crystalline axes. The latter is a process called channeled implantation and results in larger implantation depth at a given energy~\cite{Pereira_2019}.

Then, for initial proof of feasibility, we carried out a $20$~keV implantation of ${}^{229}$Th into the MgF${}_2$ resonator. ${}^{229}$Th singly charged ions were produced in an argon filled buffer gas cell as a recoil product from the alpha decay of ${}^{233}$U (source 2 in \cite{claessens_laser_2023}).  The ions were embedded in a supersonic gas jet, captured in a radiofrequency quadrupole ion guide, accelerated and mass separated with a dipole magnet \cite{ferrer_hypersonic_2021}. High-$Q$ resonators are highly sensitive to the presence of lattice defects (as those created during the implantation process), which enhance photon absorption and scattering, and thus degrade the resonator's $Q$-factor, as we observe experimentally (Fig.~\ref{fig:implantation} (c)). We estimate the implantation depth and number of defects created by an impinging ${}^{229}$Th ion at different energies in channeled (111 crystal orientation) and random orientation by simulating the creation of vacancy-interstitial pairs from the atom displacement of the impinging ion beam using MARLOWE \cite{robinson_slowing_1989} with a Frenkel pair recombination radius (capture radius for recombination of a vacancy and an interstitial) of $5.4$~\AA{}. Note that the total amount of defects created in channeled and random implantation is similar but channeled implantation requires significantly less implantation energy to achieve the same overlap with high field intensity regions. Therefore, using channeled implantation, it is possible to reach the same level of overlap at a lower implantation energy and thus lower defect density, compared to random implantation. Furthermore, defects created during implantation can be decreased using high-temperature implantation and post-implantation thermal annealing.

To gain insight into how the realistic implanted $\Th$ distributions along with spatially varying mode fields affect the nuclear transition signal, we solve the rate equations (Eq.~\ref{eq:rates}) for both random and channeled implantation techniques at various energies, taking into account the calculated electric and magnetic field distributions of the mode for the fabricated resonator and the implantation depth profile (see Fig.~\ref{fig:implantation} (b) and Fig.~\ref*{SI-fig:ImplantationDepth} (SI)). The steady state $\Phi^{\text{out}}_2$ and $\Phi_{\text{isotropic}}$ dependencies on input power at $296$~nm are shown in Fig.~\ref{fig:2photon} (b,c) for the case of $Q_1 =5\times 10^6$ (for other $Q$-factors, c.f. Fig.~\ref*{SI-fig:sandro_sweeps} in SI), with a doping concentration of $2\times10^{17}$~cm${}^{-3}$ and a technically feasible implantation fluence of $1\times10^{13}$~cm${}^{-2}$, respectively. The preliminary experimental implantation with a significantly lower fluence and electromagnetic mode overlap would require $>100$~mW input power to observe measurable cavity output photon flux at $148$~nm for the presented resonator. Nonetheless, it is clear that the resonator and implantation beam parameters can be co-optimized to maximize the overlap factor between the electromagnetic mode and the nuclear density distribution, and increase the $Q$-factor of the final implanted resonator. Implantation fluences of ${}^{229}$Th as high as $10^{13}$~cm${}^{-2}$ have already been demonstrated and a dedicated implantation station would allow for even higher fluences, resulting in steady state photon fluxes comparable to those achieved via crystal doping.

\section{\label{sec:architecture}A technology roadmap towards all solid-state nuclear clocks}

Having established the physical principles and resonator platform enabling coupling between confined optical fields and the $^{229}$Th nuclear transition, we now discuss the key technological components required to realize a practical chip-scale device. In particular, we outline realistic pathways for implementing compact laser excitation and integrated detection compatible with a scalable nanophotonic architecture.

\subsection{On-chip laser system}

One of the key challenges in realizing a scalable optical frequency reference based on the $^{229}$Th nuclear transition is the availability of a tunable, narrow-linewidth continuous-wave (CW) laser source for resonant interrogation. 
At the transition wavelength ($148.4$~nm), no commercial laser systems exist, and the available systems are limited to a small number of experimental demonstrations. 
To the best of our knowledge, only two CW sources near $148.4$~nm have been reported, a system based on second-harmonic generation in randomly quasi-phase-matched strontium tetraborate crystal (SrB$_4$O$_7$), yielding~$1$~nW of VUV power \cite{lal2025continuous, morawetz_2026}, and a cadmium-vapour four-wave-mixing source producing~$100$~nW with a narrow linewidth and tunability \cite{xiao2025continuous}.

The resonator described in the previous sections could enable efficient two-photon excitation, making it possible to use a $296.8$~nm wavelength laser.
This greatly relaxes the constraints on the technology needed to realize a suitable laser system, as lasers at this wavelength are commercially available as tabletop systems.
These systems are usually based on external-cavity laser diodes with suitable free-space frequency-doubling stages~\cite{toptica_ta_fhg_pro_2026}.
However, for a truly scalable approach, a compact or chip-scale laser system would be needed, as it would allow for the creation of a frequency reference with a footprint suitable for use outside specialized laboratory settings.
For example, laser diodes can be directly integrated with on-chip nanophotonic waveguides and resonators for linewidth narrowing and stabilization, achieving free-running linewidths at the hertz level~\cite{heim2025hybrid} in a compact form factor.
Moreover, such chip-scale laser systems have been demonstrated to interrogate the narrow-linewidth transition of $^{88}\mathrm{Sr}^+$ and to showcase the operation of a Sr-ion clock~\cite{loh2025optical}.
However, such demonstrations of on-chip narrow-linewidth lasers have been limited to the infrared and visible wavelength ranges and have not been demonstrated in the UV.
Furthermore, laser diodes at $296.8$~nm are not readily available due to challenges related to the epitaxy of the material and are limited to proof-of-concept demonstrations in research laboratories~\cite{iwaya2022recent}.
\begin{figure*}[t] 
    \centering
    \includegraphics[width=\textwidth]{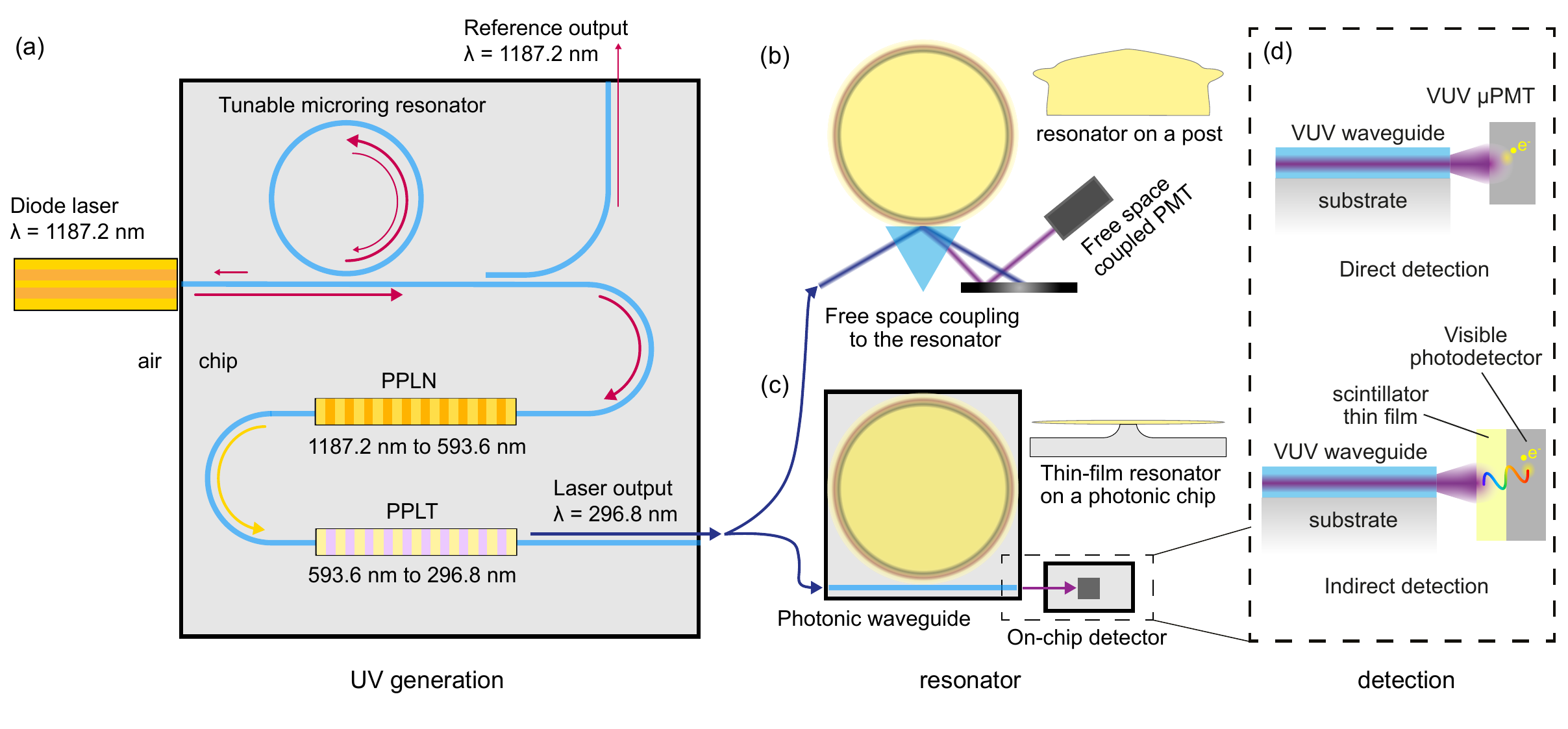}
    \caption{
    \textbf{Roadmap towards miniaturization and photonic integration of a thorium-doped resonator frequency reference.}
    (a) Schematic for an on-chip tunable laser with an excitation output wavelength of $296.8$~nm and a clock output at wavelength $1187.2$~nm to pump devices shown in (b, c). (b) Free-space implementation of a thorium-doped resonator frequency reference excited by an on-chip laser source. (c) Design for a fully integrated thorium resonator device in thin-film CaF$_2$ or other VUV transparent thin-film photonic platform. (d) Methods of on-chip VUV-photon detection.
  }
  \label{fig:integration overview}
\end{figure*}
A promising approach to reaching wavelengths not directly accessible with laser diodes is the use of on-chip frequency doubling, which has been demonstrated in several photonic platforms. 
High-performance frequency doubling has been demonstrated in periodically poled lithium niobate (PPLN) nanophotonic waveguides~\cite{xin2025wavelength} and periodically poled lithium tantalate (PPLT) nanophotonic waveguides enabled watt-level second-harmonic output at $775$~nm \cite{kuznetsov2025watt}. 
Frequency doubling in PPLT bulk waveguides has been shown with output wavelengths as short as $325$~nm and has theoretical transparency down to $280$~nm \cite{meyn1997tunable}, making it a promising candidate for frequency doubling to the two-photon excitation wavelength of the $^{229}$Th nuclear transition. 
Frequency doubling in nanophotonic aluminum nitride (AlN) waveguides has also shown high conversion efficiencies \cite{liu2023aluminum} and preliminary demonstrations down to $229$~nm~\cite{honda2023229}. 

These recent breakthroughs enable a possible pathway toward a tunable, narrow-linewidth CW laser at $296.8$~nm using only integrated components. 
A possible implementation is shown in Fig.~\ref{fig:integration overview}~(a), where a diode laser with a wavelength of $1187.2$~nm is butt-coupled to a photonic integrated chip containing a microring resonator allowing for linewidth narrowing and stabilizing of the diode laser. This is then followed by frequency doubling from $1187$~nm to $593.6$~nm using a PPLN nanophotonic waveguide and final frequency doubling using a nanophotonic PPLT waveguide from $593.6$~nm to $296.8$~nm wavelength. 

Output powers up to $4$~mW have already been demonstrated at $390$~nm using PPLN \cite{franken2025milliwatt}, and both thin-film PPLN and lithium tantalate platforms have been shown to support transfer printing, opening a pathway to fully integrated photonic systems.
Such a device has not been demonstrated at this time, but it lies within the technological capabilities of photonic integration and would allow for a mass-manufacturable and compact laser system to interrogate a thorium optical frequency standard.

The output from the chip-scale laser can then be coupled to the resonator using free-space optics, as shown in Fig.~\ref{fig:integration overview}~(b), where the enhancement of the resonator allows interrogation of the thorium transition with milliwatt level output from the chip-scale laser. Alternatively, Fig.~\ref{fig:integration overview}~(c) shows a fully-integrated device where the fluoride resonator is realized in a thin-film photonic platform.
For example, wafer-scale fabrication of thin-film CaF$_2$ was recently demonstrated \cite{song2024wafer}, and could be an interesting platform to realize thorium nanophotonics by implanting CaF$_2$ with thorium.

\subsection{Vacuum-ultraviolet photon detection}

A suitable integrated detection mechanism is needed for the proper operation of the envisioned nuclear clock. There has been growing interest in realizing on-chip detection of VUV photons, driven by research in star exploration, lithography technologies, and high-energy physics. Recent advances have focused on the miniaturization of photocathodes, wide bandgap semiconductors, and scintillator materials~\cite{zheng2020vacuum}.

Most VUV photon detectors rely on measurement of photoelectrons generated by photomultiplier tubes or wide bandgap semiconductors. Notably, materials such as aluminum nitride (AlN), aluminum gallium nitride (AlGaN), diamond, and hexagonal boron nitride (h-BN) have gained significant attention as they offer wide bandgaps, high thermal stability, and intrinsic spectral sensitivity for VUV photon detection, particularly in harsh environments such as space and synchrotrons~\cite{zheng2018vacuum2, zheng2020vacuum}. Integrated VUV semiconductor detectors have been demonstrated in AlN and AlGaN platforms~\cite{cai2021progress, jia2020vacuum, zheng2018vacuum3}. 
They can be integrated into focal plane arrays, providing micrometer spatial resolution~\cite{malinowski2011algan, zheng2018vacuum1}. 

However, photomultiplier tubes are the preferred solution for the detection of very weak VUV signals due to their ultra-high sensitivity and ultra-low noise, made possible by advancements in magnesium fluoride ($\text{MgF}_2$) window materials~\cite{paredes2018response}. Despite their superior performance, the miniaturization of PMTs for VUV detection remains a significant challenge. Traditional PMTs are inherently bulky due to the need for a vacuum enclosure and electron optics for focusing and amplification, which impedes their integration into compact systems. The development of microchannel plate PMTs ($\mu$PMTs) has enabled significant size reduction by incorporating electron multiplication within microchannels~\cite{HamamatsuMCPPMT, siegmund2018development}. Given the relatively small size of existing $\mu$PMTs, a similar approach as demonstrated in \cite{gualandi2025laser}, where a glass waveguide photonic circuit is directly coupled to a single-photon avalanche diode array, can be envisioned. In such a configuration, emitted VUV photons collected in a VUV transparent waveguide are directly coupled to a MCP from an edge facet as shown in Fig.~\ref{fig:integration overview}(d).

Another avenue for on-chip detection of VUV photons is the use of indirect scintillator detectors. Scintillator materials such as zinc oxide (ZnO) and cerium-doped yttrium aluminum garnet (YAG:Ce) have demonstrated capabilities for VUV detection, with ZnO offering ultrafast response times through doping and nanostructuring~\cite{chen2005structural, shimizu2011response}, and YAG:Ce providing high light yield and thermal stability, making it suitable for VUV imaging~\cite{dong2006luminescence}. Scintillator materials such as YAG:Ce can be integrated into nanophotonic devices, with potential in further improving their efficiency, spatial resolution, and lifetime~\cite{roques2022framework, shultzman2023enhanced, singh2024bright, kurman2024purcell, ye2024nanoplasmonic}. 
A possible implementation of a thin-film scintillator bonded to a single-photon avalanche diode is shown in Fig.~\ref{fig:integration overview}(d) and was demonstrated previously for X-ray~imaging~\cite{Wu2025CMOSSPADXray}.

\section{\label{sec:outlook}Discussion and Outlook}

Taken together, our nuclei-photon interaction analysis, proof-of-concept fabrication of $^{229}$Th-implanted WGM resonators, and the technological roadmap (compiled into Fig.~\ref{fig:integration overview}) indicate that the key performance targets for an all-solid-state nuclear clock are within reach of near-term photonic integration. Our model shows that pump powers in the milliwatt range, combined with fluoride resonators with high-$Q$ ($\sim 10^6$), nuclear doping concentrations on the order of $10^{17} \text{ } \mathrm{cm^{-3}}$ or alternatively high fluence implantation, and laser linewidths in the kHz range are sufficient to yield detectable photon fluxes on the nuclear transition using two-photon pumping in solid-state devices. 

As a first step towards realizing this roadmap, to ensure detectable photon emission rates from the nuclear transition one could collect the isotropic spontaneous emission signal, while optimized integrated designs will enable the detection of outcoupled radiation from the resonator mode. The proposed photonic platform should be~explicitly engineered to meet these specifications: high-$Q$ CaF$_2$/MgF$_2$ WGM resonators with tailored mode volumes and implantation profiles to maximize the overlap with $^{229}$Th nuclei; a fully integrated, frequency-quadrupled on-chip laser to provide tunable, narrow-linewidth light at $296.8$~nm for two-photon ONQ excitation; and compact VUV detection schemes to enable efficient readout of the $148.4$~nm photons. In this way, these building blocks chart a realistic route toward a compact, chip-scale nuclear frequency standard.

Future work will also explore another approach for enhancing clock operation -- engineering the phase-matching between the modes or implanting periodic spatial distributions of $\Th$ along the WGM resonator to enable coherent buildups and collective (superradiant) effects that can significantly impact the performance of the nuclear clock \cite{Kazakov2022}. 

Thus, the realization of the on-chip tunable laser system shown in Fig.~\ref{fig:integration overview} would bring the last major building block necessary for the implementation of our roadmap in integrated form. As other key elements for optical clocks, such as on-chip optical frequency combs and photonic microwave generation, have already been demonstrated using chip-scale components~\cite{newman_architecture_2019}, our work further motivates these developments.

\section*{Competing interests}
SK, TS, PDH, LMCP, KvG, and CRC are seeking patent protection for ideas in this work (application number EP25227524.3).

\section*{Data and code availability statement}
The data and codes that support the plots within this paper and other findings of this study are available from the corresponding authors upon reasonable request. 

\section*{Acknowledgments}
The authors acknowledge discussions with Haowei Xu and Ju Li. KM and CRC acknowledge discussions with Chen Mechel, Aliaksei Horlach, Aviv Karnieli, Swadheen~Dubey, Kjeld Beeks, Ido Kaminer, and Jamison Sloan. PDH acknowledges discussions with Takasumi Tanabe on the resonator fabrication. We thank Dennis Renisch for support in $^{233}$U source production.

\section*{Funding}
SK is supported by a fellowship from Research Foundation Flanders (FWO) under grant agreement no. 12A1026N. CRC is supported by a Stanford Science Fellowship and acknowledges startup funding from the Institute of Science and Technology, Austria (ISTA). 

Part of this work has been funded by the European Research Council (ERC) under the European Union’s Horizon 2020 research and innovation programme (Grant Agreement No. 856415), Research Foundation
Flanders (FWO, Belgium, nr. G078624N), FWO-FNRS (Belgium) under the Excellence of Science program
(Grant nr. 40007501) and the Austrian Science Fund (FWF) [Grant DOI: 10.55776/F1004, 10.55776/J4834, 10.55776/ PIN9526523]. The project 23FUN03 HIOC [Grant DOI: 10.13039/100019599] has received funding from the European Partnership on Metrology, co-financed from the European Union’s Horizon Europe Research and Innovation Program and by the Participating States.
This work was supported by the European Research Council (ERC Starting Grant LASIQ, Grant Agreement No. 101078281).

\bibliography{bibliography}

@article{brewer_27mathrm_2019,
  title   = {{$^{27}\mathrm{Al}^{+}$ Quantum-logic clock with a systematic uncertainty below $10^{-18}$}},
	volume = {123},
	pages = {033201},
	number = {3},
	journal = {Physical Review Letters},
	author = {Brewer, S. M. and Chen, J.-S. and Hankin, A. M. and Clements, E. R. and Chou, C. W. and Wineland, D. J. and Hume, D. B. and Leibrandt, D.R.},
    year = {2019}
}

@article{Savchenkov2007,
   author = {Anatoliy A Savchenkov and Andrey B Matsko and Vladimir S Ilchenko and Lute Maleki},
   issue = {11},
   journal = {Optics Express},
   month = {5},
   pages = {6768-6773},
   publisher = {Optica Publishing Group},
   title = {Optical resonators with ten million finesse},
   volume = {15},
   year = {2007},
}

@article{aeppli_clock_2024,
    title = {Clock with $8\times10^{-19}$ Systematic Uncertainty},
    volume = {133},
	pages = {023401},
	number = {2},
	journal = {Physical Review Letters},
	author = {Aeppli, Alexander and Kim, Kyungtae and Warfield, William and Safronova, Marianna S. and Ye, Jun},
	date = {2024-07-10},
    year = {2024}
	}

@article{kroger_features_1976,
  title = {Features of the low-energy level scheme of \textsuperscript{229}{T}h as observed in the $\alpha$-decay of \textsuperscript{233}{U}},
  author = {Kroger, L. A. and Reich, C. W.},
  journal = {Nuclear Physics A},
  volume = {259},
  number = {1},
  pages = {29--60},
  year = {1976},
  month = {mar},
}

@article{peik_nuclear_2003,
  title = {Nuclear laser spectroscopy of the 3.5 {eV} transition in \textsuperscript{229}{T}h},
  author = {Peik, E. and Tamm, Chr},
  journal = {Europhysics Letters},
  volume = {61},
  number = {2},
  pages = {181},
  year = {2003},
  month = {jan},
}

@article{campbell_single-ion_2012,
  title = {Single-Ion Nuclear Clock for Metrology at the 19th Decimal Place},
  author = {Campbell, C. J. and Radnaev, A. G. and Kuzmich, A. and Dzuba, V. A. and Flambaum, V. V. and Derevianko, A.},
  journal = {Physical Review Letters},
  volume = {108},
  number = {12},
  pages = {120802},
  year = {2012},
  month = {mar},
}

@article{pimon_dft_2020,
  title = {{DFT} calculation of \textsuperscript{229}{T}horium-doped magnesium fluoride for nuclear laser spectroscopy},
  author = {Pimon, M. and Gugler, J. and Mohn, P. and Kazakov, G. A. and Mauser, N. and Schumm, T.},
  journal = {Journal of Physics: Condensed Matter},
  volume = {32},
  number = {25},
  pages = {255503},
  year = {2020},
  month = {apr},
}

@article{dessovic_229thorium-doped_2014,
  title = {\textsuperscript{229}{T}horium-doped calcium fluoride for nuclear laser spectroscopy},
  author = {Dessovic, P. and Mohn, P. and Jackson, R. A. and Winkler, G. and Schreitl, M. and Kazakov, G. and Schumm, T.},
  journal = {Journal of Physics: Condensed Matter},
  volume = {26},
  number = {10},
  pages = {105402},
  year = {2014},
  month = {mar},
}

@article{kazakov_performance_2012,
  title = {Performance of a \textsuperscript{229}{T}horium solid-state nuclear clock},
  author = {Kazakov, G. A. and Litvinov, A. N. and Romanenko, V. I. and Yatsenko, L. P. and Romanenko, A. V. and Schreitl, M. and Winkler, G. and Schumm, T.},
  journal = {New Journal of Physics},
  volume = {14},
  number = {8},
  pages = {083019},
  year = {2012},
  month = {aug},
}

@article{Romanenko12,
author = {Romanenko, V. I. and Udovitskaya, Ye. G. and Yatsenko, L. P. and Romanenko, A. V. and Litvinov, A. N. and Kazakov, G. A.},
year = {2012},
title = {Direct two-photon excitation of isomeric transition in {T}horium-229 nucleus},
journal = {Ukrainian Journal of Physics},
volume = {57},
issue = {11},
pages = {1119-1131},
}

@article{rellergert_constraining_2010,
  title = {Constraining the Evolution of the Fundamental Constants with a Solid-State Optical Frequency Reference Based on the \textsuperscript{229}{T}h Nucleus},
  author = {Rellergert, Wade G. and DeMille, D. and Greco, R. R. and Hehlen, M. P. and Torgerson, J. R. and Hudson, Eric R.},
  journal = {Physical Review Letters},
  volume = {104},
  number = {20},
  pages = {200802},
  year = {2010},
  month = {may},
}

@article{von_der_wense_theory_2020,
  title = {The theory of direct laser excitation of nuclear transitions},
  author = {von der Wense, Lars and Bilous, Pavlo V. and Seiferle, Benedict and Stellmer, Simon and Weitenberg, Johannes and Thirolf, Peter G. and Pálffy, Adriana and Kazakov, Georgy},
  journal = {The European Physical Journal A},
  volume = {56},
  number = {7},
  pages = {176},
  year = {2020},
  month = {jul},
  day = {6},
}

@article{xu_solid-state_2023,
  title = {Solid-state \textsuperscript{229}{T}h nuclear laser with two-photon pumping},
  author = {Xu, Haowei and Tang, Hao and Wang, Guoqing and Li, Changhao and Li, Boning and Cappellaro, Paola and Li, Ju},
  journal = {Physical Review A},
  volume = {108},
  number = {2},
  pages = {L021502},
  year = {2023},
  month = {aug},
  day = {24},
}

@article{morgan_229th-doped_2024,
   author = {H. W.T. Morgan and R. Elwell and J. E.S. Terhune and H. B. Tran Tan and U. C. Perera and A. Derevianko and A. N. Alexandrova and E. R. Hudson},
   issue = {11},
   journal = {Applied Physics Letters},
   month = {3},
   pages = {53},
   publisher = {American Institute of Physics},
title = {Proposal and theoretical investigation of $^{229}${T}h-doped nonlinear optical crystals for compact solid-state clocks},   volume = {126},
   year = {2025},
}

@article{newman_architecture_2019,
  title = {Architecture for the photonic integration of an optical atomic clock},
  author = {Newman, Zachary L. and Maurice, Vincent and Drake, Tara and Stone, Jordan R. and Briles, Travis C. and Spencer, Daryl T. and Fredrick, Connor and Li, Qing and Westly, Daron and Ilic, B. R. and Shen, Boqiang and Suh, Myoung-Gyun and Yang, Ki Youl and Johnson, Cort and Johnson, David M. S. and Hollberg, Leo and Vahala, Kerry J. and Srinivasan, Kartik and Diddams, Scott A. and Kitching, John and Papp, Scott B. and Hummon, Matthew T.},
  journal = {Optica},
  volume = {6},
  number = {5},
  pages = {680--685},
  year = {2019},
  month = {may},
  day = {20},
}

@article{zheng2020vacuum,
  title={Vacuum-ultraviolet photon detections},
  author={Zheng, Wei and Jia, Lemin and Huang, Feng},
  journal={{i}{S}cience},
  volume={23},
  number={6},
  year={2020},
  publisher={Elsevier}
}

@article{zheng2018vacuum3,
  title={Vacuum-ultraviolet photovoltaic detector},
  author={Zheng, Wei and Lin, Richeng and Ran, Junxue and Zhang, Zhaojun and Ji, Xu and Huang, Feng},
  journal={ACS Nano},
  volume={12},
  number={1},
  pages={425--431},
  year={2018},
  publisher={ACS Publications}
}

@article{cai2021progress,
  title={Progress on {A}l{G}a{N}-based solar-blind ultraviolet photodetectors and focal plane arrays},
  author={Cai, Qing and You, Haifan and Guo, Hui and Wang, Jin and Liu, Bin and Xie, Zili and Chen, Dunjun and Lu, Hai and Zheng, Youdou and Zhang, Rong},
  journal={Light: Science \& Applications},
  volume={10},
  number={1},
  pages={94},
  year={2021},
  publisher={Nature Publishing Group UK London}
}

@article{jia2020vacuum,
  title={Vacuum-ultraviolet photodetectors},
  author={Jia, Lemin and Zheng, Wei and Huang, Feng},
  journal={PhotoniX},
  volume={1},
  pages={1--25},
  year={2020},
  publisher={Springer}
}

@article{dong2006luminescence,
  title={Luminescence studies of {C}e:{YAG} using vacuum ultraviolet synchrotron radiation},
  author={Dong, Yongjun and Zhou, Guoqing and Jun, Xu and Zhao, Guangjun and Su, Fenglian and Su, Liangbi and Zhang, Guobin and Zhang, Danhong and Li, Hongjun and Si, JiLiang},
  journal={Materials {R}esearch {B}ulletin},
  volume={41},
  number={10},
  pages={1959--1963},
  year={2006},
  publisher={Elsevier}
}

@article{chen2005structural,
  title={Structural and Optical Properties of Uniform {Z}n{O} Nanosheets},
  author={Chen, SJ and Liu, YC and Shao, CL and Mu, R and Lu, YM and Zhang, JY and Shen, DZ and Fan, XW},
  journal={Advanced Materials},
  volume={17},
  number={5},
  pages={586--590},
  year={2005}
}

@article{shimizu2011response,
  title={Response time-shortened zinc oxide scintillator for accurate single-shot synchronization of extreme ultraviolet free-electron laser and short-pulse laser},
  author={Shimizu, Toshihiko and Yamanoi, Kohei and Sakai, Kohei and Cadatal-Raduban, Marilou and Nakazato, Tomoharu and Sarukura, Nobuhiko and Kano, Masataka and Wakamiya, Akira and Ehrentraut, Dirk and Fukuda, Tsuguo and others},
  journal={Applied Physics Express},
  volume={4},
  number={6},
  pages={062701},
  year={2011},
  publisher={IOP Publishing}
}

@article{roques2022framework,
  title={A framework for scintillation in nanophotonics},
  author={Roques-Carmes, Charles and Rivera, Nicholas and Ghorashi, Ali and Kooi, Steven E and Yang, Yi and Lin, Zin and Beroz, Justin and Massuda, Aviram and Sloan, Jamison and Romeo, Nicolas and others},
  journal={Science},
  volume={375},
  number={6583},
  pages={eabm9293},
  year={2022},
  publisher={American Association for the Advancement of Science}
}

@article{shultzman2023enhanced,
  title={Enhanced imaging using inverse design of nanophotonic scintillators},
  author={Shultzman, Avner and Segal, Ohad and Kurman, Yaniv and Roques-Carmes, Charles and Kaminer, Ido},
  journal={Advanced Optical Materials},
  volume={11},
  number={8},
  pages={2202318},
  year={2023},
  publisher={Wiley Online Library}
}

@article{ye2024nanoplasmonic,
  title={The nanoplasmonic {P}urcell effect in ultrafast and high-light-yield perovskite scintillators},
  author={Ye, Wenzheng and Yong, Zhihua and Go, Michael and Kowal, Dominik and Maddalena, Francesco and Tjahjana, Liliana and Wang, Hong and Arramel, Arramel and Dujardin, Christophe and Birowosuto, Muhammad Danang and others},
  journal={Advanced Materials},
  volume={36},
  number={25},
  pages={2309410},
  year={2024},
  publisher={Wiley Online Library}
}

@article{claessens_laser_2023,
  title={Laser ionization scheme development for in-gas-jet spectroscopy studies of Th+},
  author={Claessens, A and Ivandikov, F and Bara, S and Chhetri, P and Dragoun, A and {Ch. E. Düllmann} and Elskens, Y and Ferrer, R and Kraemer, Sandro and Kudryavtsev, Yu and others},
  journal={Nuclear Instruments and Methods in Physics Research Section B: Beam Interactions with Materials and Atoms},
  volume={540},
  pages={224--226},
  year={2023},
  publisher={Elsevier}
}

@article{singh2024bright,
  title={Bright innovations: Review of next-generation advances in scintillator engineering},
  author={Singh, Pallavi and Dosovitskiy, Georgy and Bekenstein, Yehonadav},
  journal={ACS Nano},
  volume={18},
  number={22},
  pages={14029--14049},
  year={2024},
  publisher={ACS Publications}
}

@article{kurman2024purcell,
  title={Purcell-enhanced {X}-ray scintillation},
  author={Kurman, Yaniv and Lahav, Neta and Schuetz, Roman and Shultzman, Avner and Roques-Carmes, Charles and Lifshits, Alon and Zaken, Segev and Lenkiewicz, Tom and Strassberg, Rotem and Be’er, Orr and others},
  journal={Science Advances},
  volume={10},
  number={44},
  pages={eadq6325},
  year={2024},
  publisher={American Association for the Advancement of Science}
}

@article{malinowski2011algan,
  title={Al{G}a{N}-on-{S}i-based 10-$\mu$m pixel-to-pixel pitch hybrid imagers for the {EUV} range},
  author={Malinowski, Pawel E and Duboz, Jean-Yves and De Moor, Piet and John, Joachim and Minoglou, Kyriaki and Srivastava, Puneet and Semond, Fabrice and Frayssinet, Eric and Giordanengo, Boris and Ben Moussa, Ali and others},
  journal={IEEE {E}lectron {D}evice {L}etters},
  volume={32},
  number={11},
  pages={1561--1563},
  year={2011},
  publisher={IEEE}
}

@article{paredes2018response,
  title={Response of photomultiplier tubes to xenon scintillation light},
  author={Paredes, B L{\'o}pez and Ara{\'u}jo, HM and Froborg, F and Marangou, N and Olcina, I and Sumner, TJ and Taylor, R and Tom{\'a}s, A and Vacheret, A},
  journal={Astroparticle Physics},
  volume={102},
  pages={56--66},
  year={2018},
  publisher={Elsevier}
}

@article{zheng2018vacuum1,
  title={Vacuum ultraviolet photovoltaic arrays},
  author={Zheng, Wei and Lin, Richeng and Jia, Lemin and Huang, Feng},
  journal={Photonics Research},
  volume={7},
  number={1},
  pages={98--102},
  year={2018},
  publisher={Chinese Laser Press and Optical Society of America}
}

@misc{HamamatsuMCPPMT,
  author       = {{Hamamatsu Photonics}},
  title        = {{MCP-PMT} - {M}icrochannel {P}late {P}hotomultiplier {T}ube},
  year         = {2025},
  url          = {https://www.hamamatsu.com/us/en/product/optical-sensors/pmt/pmt_tube-alone/mcp-pmt.html},
  note         = {Accessed: 2025-02-10}
}

@inproceedings{siegmund2018development,
  title={Development of sealed tube microchannel plate detectors with cross strip readouts},
  author={Siegmund, O and Ertley, C and Darling, N and Curtis, T and Hull, J and Vallerga, J and Paw, CU},
  booktitle={2018 IEEE Nuclear Science Symposium and Medical Imaging Conference Proceedings (NSS/MIC)},
  pages={1--7},
  year={2018},
  organization={IEEE}
}

@article{zheng2018vacuum2,
  title={Vacuum-ultraviolet photodetection in few-layered h-{BN}},
  author={Zheng, Wei and Lin, Richeng and Zhang, Zhaojun and Huang, Feng},
  journal={ACS Applied Materials \& Interfaces},
  volume={10},
  number={32},
  pages={27116--27123},
  year={2018},
  publisher={ACS Publications}
}

@article{helmer_excited_1994,
  title={An excited state of {T}h-229 at 3.5 e{V}},
  author={Helmer, RG and Reich, CW},
  journal={Physical Review C},
  volume={49},
  number={4},
  pages={1845},
  year={1994},
  publisher={APS}
}

@article{beck_energy_2007,
  title = {Energy splitting of the ground-state doublet in the nucleus Th 229},
  author = {Beck, B. R. and Becker, J. A. and Beiersdorfer, P. and Brown, G. V. and Moody, Kenton J. and Wilhelmy, Jerry B. and Porter, F. S. and Kilbourne, C. A. and Kelley, R. L.},
  journal = {Physical Review Letters},
  volume = {98},
  number = {14},
  pages = {142501},
  year = {2007}
}

@techreport{beck_improved_2009,
  title={Improved value for the energy splitting of the ground-state doublet in the nucleus 229mTh},
  author={Beck, BR and Wu, C and Beiersdorfer, P and Brown, GV and Becker, JA and Moody, KJ and Wilhelmy, JB and Porter, FS and Kilbourne, CA and Kelley, RL},
  year={2009},
  institution={Lawrence Livermore National Laboratory (LLNL), Livermore, CA}
}

@article{von_der_wense_direct_2016,
  title = {Direct detection of the $^{229}${T}h nuclear clock transition},
  author = {von der Wense, Lars and Seiferle, Benedict and Laatiaoui, Mustapha and Neumayr, Jürgen B. and Maier, Hans-Jörg and Wirth, Hans-Friedrich and Mokry, Christoph and Runke, Jörg and Eberhardt, Klaus and {Ch. E. Düllmann} and Trautmann, Norbert G. and Thirolf, Peter G.},
  journal = {Nature},
  volume = {533},
  number = {7601},
  pages = {47--51},
  year = {2016},
  month = {may},
}

@article{thielking_laser_2018,
  title = {Laser spectroscopic characterization of the nuclear-clock isomer \textsuperscript{229m}{Th}},
  author = {Thielking, Johannes and Okhapkin, Maxim V. and Głowacki, Przemysław and Meier, David M. and von der Wense, Lars and Seiferle, Benedict and {Ch. E. Düllmann}. and Thirolf, Peter G. and Peik, Ekkehard},
  journal = {Nature},
  volume = {556},
  number = {7701},
  pages = {321--325},
  year = {2018},
  month = {apr}
}

@article{seiferle_lifetime_2017,
  title = {Lifetime Measurement of the \textsuperscript{229m}{Th} Nuclear Isomer},
  author = {Seiferle, Benedict and von der Wense, Lars and Thirolf, Peter G.},
  journal = {Physical Review Letters},
  volume = {118},
  number = {4},
  pages = {042501},
  year = {2017},
  month = {jan},
  day = {26},
}

@article{seiferle_energy_2019,
  title = {Energy of the \textsuperscript{229}{Th} Nuclear Clock Transition},
  author = {Seiferle, Benedict and von der Wense, Lars and Bilous, Pavlo V. and Amersdorffer, Ines and Lemell, Christoph and Libisch, Florian and Stellmer, Simon and Schumm, Thorsten and {Ch. E. Düllmann} and Pálffy, Adriana and Thirolf, Peter G.},
  journal = {Nature},
  volume = {573},
  number = {7773},
  pages = {243--246},
  year = {2019},
  month = {sep},
}

@article{verlinde_alternative_2019,
  title={Alternative approach to populate and study the Th-229 nuclear clock isomer},
  author={Verlinde, M and Kraemer, S and Moens, J and Chrysalidis, K and Correia, JG and Cottenier, Stefaan and De Witte, H and Fedorov, DV and Fedosseev, VN and Ferrer, R and others},
  journal={Physical Review C},
  volume={100},
  number={2},
  pages={024315},
  year={2019},
  publisher={APS}
}

@article{kraemer_observation_2023,
	title = {Observation of the radiative decay of the $^{229}${T}h nuclear clock isomer},
	volume = {617},
	pages = {706--710},
	number = {7962},
	journal = {Nature},
	author = {Kraemer, Sandro and Moens, Janni and Athanasakis-Kaklamanakis, Michail and Bara, Silvia and Beeks, Kjeld and Chhetri, Premaditya and Chrysalidis, Katerina and Claessens, Arno and Cocolios, Thomas E. and Correia, João G. M. and Witte, Hilde De and Ferrer, Rafael and Geldhof, Sarina and Heinke, Reinhard and Hosseini, Niyusha and Huyse, Mark and Köster, Ulli and Kudryavtsev, Yuri and Laatiaoui, Mustapha and Lica, Razvan and Magchiels, Goele and Manea, Vladimir and Merckling, Clement and Pereira, Lino M. C. and Raeder, Sebastian and Schumm, Thorsten and Sels, Simon and Thirolf, Peter G. and Tunhuma, Shandirai Malven and Van Den Bergh, Paul and Van Duppen, Piet and Vantomme, André and Verlinde, Matthias and Villarreal, Renan and Wahl, Ulrich},
	date = {2023-05},
    year = {2023},
	}

@article{zhang_frequency_2024,
title = {Frequency ratio of the $^{229m}${T}h nuclear isomeric transition and the $^{87}${S}r atomic clock},	volume = {633},
	pages = {63--70},
	number = {8028},
	journal = {Nature},
	author = {Zhang, Chuankun and Ooi, Tian and Higgins, Jacob S. and Doyle, Jack F. and von der Wense, Lars and Beeks, Kjeld and Leitner, Adrian and Kazakov, Georgy A. and Li, Peng and Thirolf, Peter G. and Schumm, Thorsten and Ye, Jun},
	date = {2024-09},
    year = {2024}
}

@article{elwell_laser_2024,
	title = {Laser Excitation of the \textsuperscript{229}{T}h Nuclear Isomeric Transition in a Solid-State Host},
	volume = {133},
	pages = {013201},
	number = {1},
	journal = {Physical Review Letters},
	shortjournal = {Phys. Rev. Lett.},
	author = {Elwell, R. and Schneider, Christian and Jeet, Justin and Terhune, J. E. S. and Morgan, H. W. T. and Alexandrova, A.N. and Tran Tan, H. B. and Derevianko, Andrei and Hudson, Eric R.},
	year = {2024}
}

@article{tiedau_laser_2024,
	title = {Laser Excitation of the {T}h-229 Nucleus},
	volume = {132},
	pages = {182501},
	number = {18},
	journal = {Physical Review Letters},
	shortjournal = {Phys. Rev. Lett.},
	author = {Tiedau, J. and Okhapkin, M. V. and Zhang, K. and Thielking, J. and Zitzer, G. and Peik, E. and Schaden, F. and Pronebner, T. and Morawetz, I. and De Col, L. Toscani and Schneider, F. and Leitner, A. and Pressler, M. and Kazakov, G. A. and Beeks, K. and Sikorsky, T. and Schumm, T.},
	date = {2024-04-29},
    year = {2024},
	}

@article{zhang_229thf4_2024,
title = {$^{229}${T}h{F}$_4$ thin films for solid-state nuclear clocks},	volume = {636},
	pages = {603--608},
	number = {8043},
	journal = {Nature},
	author = {Zhang, Chuankun and von der Wense, Lars and Doyle, Jack F. and Higgins, Jacob S. and Ooi, Tian and Friebel, Hans U. and Ye, Jun and Elwell, R. and Terhune, J. E. S. and Morgan, H. W. T. and Alexandrova, A. N. and Tran Tan, H. B. and Derevianko, Andrei and Hudson, Eric R.},
	date = {2024-12},
    year = {2024}
	}

@article{pineda_radiative_2025,
	title = {Radiative decay of the \textsuperscript{229}{T}h nuclear clock isomer in different host materials},
	volume = {7},
	pages = {013052},
	number = {1},
	journal = {Physical Review Research},
	shortjournal = {Phys. Rev. Res.},
	author = {Pineda, S. V. and Chhetri, P. and Bara, S. and Elskens, Y. and Casci, S. and Alexandrova, A. N. and Au, M. and Athanasakis-Kaklamanakis, M. and Bartokos, M. and Beeks, K. and Bernerd, C. and Claessens, A. and Chrysalidis, K. and Cocolios, T. E. and Correia, J. G. and De Witte, H. and Elwell, R. and Ferrer, R. and Heinke, R. and Hudson, E. R. and Ivandikov, F. and Kudryavtsev, Yu. and Köster, U. and Kraemer, S. and Laatiaoui, M. and Lica, R. and Merckling, C. and Morawetz, I. and Morgan, H. W. T. and Moritz, D. and Pereira, L. M. C. and Raeder, S. and Rothe, S. and Schaden, F. and Scharl, K. and Schumm, T. and Stegemann, S. and Terhune, J. and Thirolf, P. G. and Tunhuma, S. M. and Van Den Bergh, P. and Van Duppen, P. and Vantomme, A. and Wahl, U. and Yue, Z.},
    year = {2025}
	}

@article{higgins_temperature_2025,
  title={Temperature sensitivity of a {T}horium-229 solid-state nuclear clock},
  author={Higgins, Jacob S and Ooi, Tian and Doyle, Jack F and Zhang, Chuankun and Ye, Jun and Beeks, Kjeld and Sikorsky, Tomas and Schumm, Thorsten},
  journal={Physical Review Letters},
  volume={134},
  number={11},
  pages={113801},
  year={2025},
  publisher={APS}
}

@article{minkov_reduced_2017,
	title = {Reduced Transition Probabilities for the Gamma Decay of the 7.8 {eV} Isomer in \textsuperscript{229}{T}h},
	volume = {118},
	pages = {212501},
	number = {21},
	journal = {Physical Review Letters},
	shortjournal = {Phys. Rev. Lett.},
	author = {Minkov, Nikolay and Pálffy, Adriana},
	date = {2017-05-23},
    year = {2017}
}

@article{tkalya_radiative_2015,
title = {Radiative lifetime and energy of the low-energy isomeric level in $^{229}\mathrm{Th}$},	volume = {92},
	pages = {054324},
	number = {5},
	journal = {Physical Review C},
	author = {Tkalya, E. V. and Schneider, Christian and Jeet, Justin and Hudson, Eric R.},
	date = {2015-11-25},
    year = {2015}
}

@article{ruchowska_nuclear_2006,
	title = {Nuclear structure of $^{229}${T}h},
	volume = {73},
	pages = {044326},
	number = {4},
	journal = {Physical Review C},
	shortjournal = {Phys. Rev. C},
	author = {Ruchowska, E. and Płóciennik, W. A. and Żylicz, J. and Mach, H. and Kvasil, J. and Algora, A. and Amzal, N. and Bäck, T. and Borge, M. G. and Boutami, R. and Butler, P. A. and Cederkäll, J. and Cederwall, B. and Fogelberg, B. and Fraile, L. M. and Fynbo, H. O. U. and Hagebø, E. and Hoff, P. and Gausemel, H. and Jungclaus, A. and Kaczarowski, R. and Kerek, A. and Kurcewicz, W. and Lagergren, K. and Nacher, E. and Rubio, B. and Syntfeld, A. and Tengblad, O. and Wasilewski, A. A. and Weissman, L.},
	date = {2006-04-26},
    year = {2026}
}

@article{Anderson2018,
  author    = {M. Anderson and N. G. Pavlov and J. D. Jost and G. Lihachev and J. Liu and T. Morais and M. Zervas and M. L. Gorodetsky and T. J. Kippenberg},
  title     = {Highly efficient coupling of crystalline microresonators to integrated photonic waveguides},
  journal   = {Optics Letters},
  volume    = {43},
  number    = {9},
  pages     = {2106--2109},
  year      = {2018},
}

@article{liulowloss2018,
  title = {Low-Loss Prism-Waveguide Optical Coupling for Ultrahigh-${Q}$ Low-Index Monolithic Resonators},
  author = {Liu, Guangyao and Ilchenko, Vladimir S. and Su, Tiehui and Ling, Yi-Chun and Feng, Shaoqi and Shang, Kuanping and Zhang, Yu and Liang, Wei and Savchenkov, Anatoliy A. and Matsko, Andrey B. and Maleki, Lute and Yoo, S. J. Ben},
  year = 2018,
  journal = {Optica},
  volume = {5},
  number = {2},
  pages = {219--226},
  urldate = {2026-01-19},
  copyright = {\copyright{} 2018 Optical Society of America},
  langid = {english}
}

@article{kitching2018chip,
  title={Chip-scale atomic devices},
  author={Kitching, John},
  journal={Applied Physics Reviews},
  pages={031302},
  volume={5},
  number={3},
  year={2018},
  publisher={AIP Publishing}
}

@article{vahala2003optical,
  title={Optical microcavities},
  author={Vahala, Kerry J},
  journal={Nature},
  volume={424},
  number={6950},
  pages={839--846},
  year={2003},
  publisher={Nature Publishing Group UK London}
}

@article{ilchenko2006optical,
  title={Optical resonators with whispering-gallery modes-part {II}: applications},
  author={Ilchenko, Vladimir S and Matsko, Andrey B},
  journal={IEEE Journal of selected topics in quantum electronics},
  volume={12},
  number={1},
  pages={15--32},
  year={2006},
  publisher={IEEE}
}

@article{matsko2006optical,
  title={Optical resonators with whispering gallery modes {I}: basics},
  author={Matsko, Andrey B and Ilchenko, Vladimir S},
  journal={IEEE Journal of selected topics in quantum electronics},
  volume={12},
  number={3},
  pages={3},
  year={2006}
}

@article{yang2025systemaic,
    author = {Yang, Liu and Takabayashi, Ryomei and Moriguchi, Hiroki and Kodama, Hikaru and Miura, Kazuma and Wada, Koshiro and Yamaguchi, Kai and Murakami, Tatsuki and Kumazaki, Hajime and Kakinuma, Yasuhiro and Tanabe, Takasumi and Fujii, Shun},
    title = {Systematic dispersion engineering of crystalline microresonators for broadband and coherent frequency comb generation},
    journal = {APL Photonics},
    volume = {10},
    number = {12},
    pages = {126109},
    year = {2025},
    month = {12},
}

@article{spillane_ideality_2003,
  title={Ideality in a fiber-taper-coupled microresonator system for application to cavity quantum electrodynamics},
  author={Spillane, SM and Kippenberg, TJ and Painter, OJ and Vahala, KJ},
  journal={Physical Review Letters},
  volume={91},
  number={4},
  pages={043902},
  year={2003},
  publisher={APS}
}

@article{gorodetsky_optical_1999,
  title={Optical microsphere resonators: optimal coupling to high-${Q}$ whispering-gallery modes},
  author={Gorodetsky, Michael L and Ilchenko, Vladimir S},
  journal={Journal of the Optical Society of America B},
  volume={16},
  number={1},
  pages={147--154},
  year={1999},
  publisher={Optical Society of America}
}

@article{furst_naturally_2010,
  title={Naturally Phase-Matched Second-Harmonic Generation in a Whispering-Gallery-Mode Resonator},
  author={F{\"u}rst, JU and Strekalov, DV and Elser, Dominique and Lassen, Mikael and Andersen, Ulrik Lund and Marquardt, Christoph and Leuchs, Gerd},
  journal={Physical Review Letters},
  volume={104},
  number={15},
  pages={153901},
  year={2010},
  publisher={APS}
}

@article{strekalov_nonlinear_2016,
  title={Nonlinear and quantum optics with whispering gallery resonators},
  author={Strekalov, Dmitry V and Marquardt, Christoph and Matsko, Andrey B and Schwefel, Harald GL and Leuchs, Gerd},
  journal={Journal of Optics},
  volume={18},
  number={12},
  pages={123002},
  year={2016},
  publisher={IOP Publishing}
}

@article{grudininfundamental2007,
  title = {On the Fundamental Limits of {{Q}} Factor of Crystalline Dielectric Resonators},
  author = {Grudinin, Ivan S. and Matsko, Andrey B. and Maleki, Lute},
  year = {2007},
  journal = {Optics Express},
  shortjournal = {Opt. Express, OE},
  volume = {15},
  number = {6},
  pages = {3390--3395},
  doi = {10.1364/OE.15.003390},
  langid = {english}
}

@article{savchenkovselfinjection2019,
  title = {Self-Injection Locking Efficiency of a {{UV Fabry}}--{{Perot}} Laser Diode},
  author = {Savchenkov, Anatoliy A. and Chiow, Sheng-Wey and Ghasemkhani, Mohammadreza and Williams, Skip and Yu, Nan and Stirbl, Robert C. and Matsko, Andrey B.},
  year = 2019,
  journal = {Optics Letters},
  volume = {44},
  number = {17},
  pages = {4175--4178},
  urldate = {2026-01-19},
  copyright = {\copyright{} 2019 Optical Society of America},
  langid = {english}
}

@article{Liu2021,
   author = {Xianwen Liu and Zheng Gong and Alexander W. Bruch and Joshua B. Surya and Juanjuan Lu and Hong X. Tang},
   issue = {1},
   journal = {Nature Communications},
   month = {9},
   pages = {5428-},
   pmid = {34521858},
   publisher = {Nature Publishing Group},
   title = {Aluminum nitride nanophotonics for beyond-octave soliton microcomb generation and self-referencing},
   volume = {12},
   year = {2021},
}

@article{Wu2025,
  title={Vernier microcombs for integrated optical atomic clocks},
  author={Wu, Kaiyi and O’Malley, Nathan P and Fatema, Saleha and Wang, Cong and Girardi, Marcello and Alshaykh, Mohammed S and Ye, Zhichao and Leaird, Daniel E and Qi, Minghao and Torres-Company, Victor and others},
  journal={Nature Photonics},
  volume={19},
  number={4},
  pages={400--406},
  year={2025},
  publisher={Nature Publishing Group UK London}
}

@article{breunig_three_2016,
  title={Three-wave mixing in whispering gallery resonators},
  author={Breunig, Ingo},
  journal={Laser \& Photonics Reviews},
  volume={10},
  number={4},
  pages={569--587},
  year={2016},
  publisher={Wiley Online Library}
}

@article{Lin2015,
   author = {Jintian Lin and Yingxin Xu and Zhiwei Fang and Min Wang and Jiangxin Song and Nengwen Wang and Lingling Qiao and Wei Fang and Ya Cheng},
   issue = {1},
   journal = {Scientific Reports},
   month = {1},
   pages = {8072-},
   publisher = {Nature Publishing Group},
   title = {Fabrication of high-${Q}$ lithium niobate microresonators using femtosecond laser micromachining},
   volume = {5},
   year = {2015},
}

@article{savchenkov2015generation,
  title={Generation of {K}err combs centered at 4.5 $\mu$m in crystalline microresonators pumped with quantum-cascade lasers},
  author={Savchenkov, Anatoliy A and Ilchenko, Vladimir S and Di Teodoro, Fabio and Belden, Paul M and Lotshaw, William T and Matsko, Andrey B and Maleki, Lute},
  journal={Optics Letters},
  volume={40},
  number={15},
  pages={3468--3471},
  year={2015},
  publisher={Optical Society of America}
}

@article{grudinin2006ultra,
  title={Ultra high ${Q}$ crystalline microcavities},
  author={Grudinin, Ivan S and Matsko, Andrey B and Savchenkov, Anatoliy A and Strekalov, Dmitry and Ilchenko, Vladimir S and Maleki, Lute},
  journal={Optics Communications},
  volume={265},
  number={1},
  pages={33--38},
  year={2006},
  publisher={Elsevier}
}

@techreport{grudinin2008single,
  title={Single-Mode WGM Resonators Fabricated by Diamond Turning},
  author={Grudinin, Ivan and Maleki, Lute and Savchenkov, Anatoliy and Matsko, Andrewy and Strekalov, Dmitry and Iltchenko, Vladimir},
  year={2008},
  institution = {NASA’s Jet Propulsion Laboratory, Pasadena, California},
}

@article{lin2017nonlinear,
  title={Nonlinear photonics with high-${Q}$ whispering-gallery-mode resonators},
  author={Lin, Guoping and Coillet, Aur{\'e}lien and Chembo, Yanne K},
  journal={Advances in Optics and Photonics},
  volume={9},
  number={4},
  pages={828--890},
  year={2017},
  publisher={Optical Society of America}
}

@article{Li2023,
   author = {Lin Li and Zi Li and Chen Wang and Wen Ting Gan and Xia Hua and Xin Tong},
   issue = {2},
   journal = {Nuclear Science and Techniques},
   month = {2},
   pages = {24-},
   publisher = {Springer},
   title = {Scheme for the excitation of {T}horium-229 nuclei based on electronic bridge excitation},
   volume = {34},
   year = {2023},
}

@article{xu2023two,
  title={Two-photon interface of nuclear spins based on the optonuclear quadrupolar effect},
  author={Xu, Haowei and Li, Changhao and Wang, Guoqing and Wang, Hua and Tang, Hao and Barr, Ariel Rebekah and Cappellaro, Paola and Li, Ju},
  journal={Physical Review X},
  volume={13},
  number={1},
  pages={011017},
  year={2023},
  publisher={APS}
}

@article{gorodetsky_rayleigh_2000,
	title = {Rayleigh scattering in high-${Q}$ microspheres},
	volume = {17},
	pages = {1051--1057},
	number = {6},
	journal = {{JOSA} B},
	author = {Gorodetsky, Michael L. and Pryamikov, Andrew D. and Ilchenko, Vladimir S.},
    year = {2000}
}

@book{joannopoulos2011photonic,
  title={Photonic Crystals: Molding the Flow of Light - Second Edition},
  author={Joannopoulos, J.D. and Johnson, S.G. and Winn, J.N. and Meade, R.D.},
  year={2011},
  publisher={Princeton University Press}
}

@article{Kazakov2022,
   author = {Georgy A. Kazakov and Swadheen Dubey and Anna Bychek and Uwe Sterr and Marcin Bober and Michał Zawada},
   issue = {5},
   journal = {Physical Review A},
   month = {11},
   pages = {053114},
   publisher = {American Physical Society},
   title = {Ultimate stability of active optical frequency standards},
   volume = {106},
   year = {2022},
}

@article{fujii_all-precision-machining_2020,
	title = {All-precision-machining fabrication of ultrahigh-${Q}$ crystalline optical microresonators},
	volume = {7},
	pages = {694--701},
	number = {6},
	journal = {Optica},
	author = {Fujii, Shun and Hayama, Yuka and Imamura, Kosuke and Kumazaki, Hajime and Kakinuma, Yasuhiro and Tanabe, Takasumi},
    year = {2020}
}

@article{lal2025continuous,
  title={Continuous-wave laser source at the 148 nm nuclear transition of {T}h-229},
  author={Lal, Vishal and Okhapkin, Maksim V and Tiedau, Johannes and Irwin, Niels and Petrov, Valentin and Peik, Ekkehard},
  journal={Optica},
  volume={12},
  number={12},
  pages={1971--1974},
  year={2025},
  publisher={Optica Publishing Group}
}

@article{morawetz_2026,
  title         = {Continuous-wave nuclear laser absorption spectroscopy of {T}horium-229},
  author        = {Morawetz, I. and Riebner, T. and Toscani De Col, L. and Schneider, F. and
                   Sempelmann, N. and Schaden, F. and Bartokos, M. and Kazakov, G. A. and
                   Lahs, S. and Beeks, K. and Gerstenecker, B. and Gr{\"u}neis, A. and
                   Pimon, M. and Schumm, T. and Lal, V. and Zitzer, G. and Tiedau, J. and
                   Okhapkin, M. V. and Petrov, V. and Peik, E.},
  journal       = {arXiv preprint arXiv:2604.16640},
  year          = {2026},
  month         = apr,
}

@article{xiao2025continuous,
  title={Continuous-wave narrow-linewidth vacuum ultraviolet laser source},
  author={Xiao, Qi and Penyazkov, Gleb and Li, Xiangliang and Huang, Beichen and Bu, Wenhao and Shi, Juanlang and Shi, Haoyu and Liao, Tangyin and Yan, Gaowei and Tian, Haochen and others},
  journal={Nature},
  pages={1--5},
  year={2026},
  publisher={Nature Publishing Group UK London}
}

@misc{toptica_ta_fhg_pro_2026,
  author       = {{TOPTICA Photonics SE}},
  title        = {{TA-FHG pro: High-power, tunable, frequency-quadrupled diode laser}},
  howpublished = {\url{https://www.toptica.com/products/tunable-diode-lasers/frequency-converted-lasers/ta-fhg-pro}},
  note         = {Accessed: 2026-01-03},
  year = {2026}
}

@article{iwaya2022recent,
  title={Recent development of {UV}-{B} laser diodes},
  author={Iwaya, Motoaki and Tanaka, Shunya and Omori, Tomoya and Yamada, Kazuki and Hasegawa, Ryota and Shimokawa, Moe and Yabutani, Ayumu and Iwayama, Sho and Sato, Kosuke and Takeuchi, Tetsuya and others},
  journal={Japanese Journal of Applied Physics},
  volume={61},
  number={4},
  pages={040501},
  year={2022},
  publisher={IOP Publishing}
}

@article{kuznetsov2025watt,
  title={Watt-level second harmonic generation in periodically poled thin-film lithium tantalate},
  author={Kuznetsov, Nikolai and Li, Zihan and Kippenberg, Tobias J},
  journal={arXiv preprint arXiv:2512.07968},
  year={2025}
}

@article{loh2025optical,
  title={Optical atomic clock interrogation using an integrated spiral cavity laser},
  author={Loh, William and Reens, David and Kharas, Dave and Sumant, Alkesh and Belanger, Connor and Maxson, Ryan T and Medeiros, Alexander and Setzer, William and Gray, Dodd and DeBry, Kyle and others},
  journal={Nature Photonics},
  volume={19},
  number={3},
  pages={277--283},
  year={2025},
  publisher={Nature Publishing Group UK London}
}

@article{heim2025hybrid,
  title={Hybrid integrated ultra-low linewidth coil stabilized isolator-free widely tunable external cavity laser},
  author={Heim, David AS and Bose, Debapam and Liu, Kaikai and Isichenko, Andrei and Blumenthal, Daniel J},
  journal={Nature Communications},
  volume={16},
  number={1},
  pages={5944},
  year={2025},
  publisher={Nature Publishing Group UK London}
}

@article{xin2025wavelength,
  title={Wavelength-accurate and wafer-scale process for nonlinear frequency mixers in thin-film lithium niobate},
  author={Xin, CJ and Lu, Shengyuan and Yang, Jiayu and Shams-Ansari, Amirhassan and Desiatov, Boris and Magalh{\~a}es, Let{\'\i}cia S and Ghosh, Soumya S and McGee, Erin and Renaud, Dylan and Achuthan, Nicholas and others},
  journal={Communications Physics},
  volume={8},
  number={1},
  pages={136},
  year={2025},
  publisher={Nature Publishing Group UK London}
}

@article{honda2023229,
  title={229 nm far-ultraviolet second harmonic generation in a vertical polarity inverted AlN bilayer channel waveguide},
  author={Honda, Hiroto and Umeda, Soshi and Shojiki, Kanako and Miyake, Hideto and Ichikawa, Shuhei and Tatebayashi, Jun and Fujiwara, Yasufumi and Serita, Kazunori and Murakami, Hironaru and Tonouchi, Masayoshi and others},
  journal={Applied Physics Express},
  volume={16},
  number={6},
  pages={062006},
  year={2023},
  publisher={IOP Publishing}
}

@article{liu2023aluminum,
  title={Aluminum nitride photonic integrated circuits: from piezo-optomechanics to nonlinear optics},
  author={Liu, Xianwen and Bruch, Alexander W and Tang, Hong X},
  journal={Advances in Optics and Photonics},
  volume={15},
  number={1},
  pages={236--317},
  year={2023},
  publisher={Optica Publishing Group}
}

@article{meyn1997tunable,
  title={Tunable ultraviolet radiation by second-harmonic generation in periodically poled lithium tantalate},
  author={Meyn, J-P and Fejer, MM},
  journal={Optics Letters},
  volume={22},
  number={16},
  pages={1214--1216},
  year={1997},
  publisher={Optical Society of America}
}

@article{franken2025milliwatt,
  title={Milliwatt-level {UV} generation using sidewall poled lithium niobate},
  author={Franken, Cornelis AA and Ghosh, Soumya S and Rodrigues, Caique C and Yang, Jiayu and Xin, Chen J and Lu, Shengyuan and Witt, Donald and Joe, G and Wiederhecker, GS and Boller, K-J and others},
  journal={arXiv preprint arXiv:2503.16785},
  year={2025}
}

@article{song2024wafer,
  title={Wafer-scale fabrication of single-crystalline calcium fluoride thin-film on insulator by ion-cutting},
  author={Song, Qiudong and Cai, Jiachen and Wang, Chengli and Zhou, Liping and Chen, Yang and Zhou, Min and Zhang, Jian and Yang, Bingcheng and Yang, Yi and Yi, Ailun and others},
  journal={Optical Materials},
  volume={157},
  pages={115787},
  year={2024},
  publisher={Elsevier}
}

@article{ooi_frequency_2026,
   author = {Tian Ooi and Jack F. Doyle and Chuankun Zhang and Jacob S. Higgins and Jun Ye and Kjeld Beeks and Tomas Sikorsky and Thorsten Schumm},
   issue = {8100},
   journal = {Nature},
   month = {1},
   pages = {72-78},
   publisher = {Nature Publishing Group},
   title = {Frequency reproducibility of solid-state {T}horium-229 nuclear clocks},
   volume = {650},
   year = {2026},
}

@article{ludlow2015optical,
  title={Optical atomic clocks},
  author={Ludlow, Andrew D and Boyd, Martin M and Ye, Jun and Peik, Ekkehard and Schmidt, Piet O},
  journal={Reviews of Modern Physics},
  volume={87},
  number={2},
  pages={637--701},
  year={2015},
  publisher={APS}
}

@article{ferrer_hypersonic_2021,
  title={Hypersonic nozzle for laser-spectroscopy studies at 17 K characterized by resonance-ionization-spectroscopy-based flow mapping},
  author={Ferrer, R and Verlinde, M and Verstraelen, E and Claessens, A and Huyse, M and Kraemer, S and Kudryavtsev, Yu and Romans, J and Van den Bergh, P and Van Duppen, P and others},
  journal={Physical Review Research},
  volume={3},
  number={4},
  pages={043041},
  year={2021},
  publisher={APS}
}

@article{robinson_slowing_1989,
  title = {Slowing-down time of energetic atoms in solids},
  author = {Robinson, Mark T.},
  journal = {Physical Review B},
  volume = {40},
  issue = {16},
  pages = {10717--10726},
  numpages = {0},
  year = {1989},
  month = {Dec},
  publisher = {American Physical Society},
}

@incollection{Pereira_2019, year = 2019, month = {sep}, publisher = {Institution of Engineering and Technology}, pages = {501--561}, author = {Lino Pereira and Andre Vantomme and Ulrich Wahl}, title = {Characterizing defects with ion beam analysis and channeling techniques}, booktitle = {Characterisation and Control of Defects in Semiconductors}}

@article{gualandi2025laser,
  title={Laser-written reconfigurable photonic integrated circuit directly coupled to a single-photon avalanche diode array},
  author={Gualandi, Giulio and Atzeni, Simone and Gardina, Marco and Caime, Antonino and Corrielli, Giacomo and Labanca, Ivan and Gulinatti, Angelo and Rech, Ivan and Osellame, Roberto and Acconcia, Giulia and others},
  journal={Light: Science \& Applications},
  volume={14},
  number={1},
  pages={199},
  year={2025},
  publisher={Nature Publishing Group UK London}
}

@inproceedings{Wu2025CMOSSPADXray,
  author    = {Jau-Yang Wu and Chun-Wei Tang and Chun-Hsien Liu and Bi-Hsuan Lin and Chia-Ming Tsai and Sheng-Di Lin},
  title     = {High Timing and Spatial Resolution {X}-Ray Imaging Demonstrated by Direct Coating of Scintillator on a {CMOS} {SPAD} Array},
  booktitle = {OECC/PSC 2025},
  year      = {2025},
  pages     = {1--3},
}

\onecolumngrid
\clearpage
\newpage

\makeatletter
\let\tableofcontents\relax
\let\listoffigures\relax
\let\listoftables\relax
\let\p@appendix\relax
\makeatother

\ifarXiv
    \setcounter{page}{1}
    \pagestyle{empty}
    
    \hbox{}
    \clearpage
    
    

\section*{Supplementary material}

Contents

---

*Electronic address: sandro.kraemer@kuleuven.be; These authors contributed equally.

†Electronic address: karen.mamian@ist.ac.at; These authors contributed equally.

‡Present address: Forschungszentrum Jülich, Jülich, Germany

§Electronic address: thorsten.schumm@tuwien.ac.at

¶Electronic address: pascal.delhaye@mpl.mpg.de

**Electronic address: lino.pereira@kuleuven.be

††Electronic address: kazakov.george@gmail.com

‡‡Electronic address: kasper.vangasse@ugent.be

§§Electronic address: crc@ista.ac.at

## S1. BASIC CONCEPTS OF NANOPHOTONICS AND COUPLED-MODE THEORY

In this first section, we review the basic concepts of nanophotonics and coupled-mode theory used in this paper. Nanophotonic cavities support a set of modes that can be described by bosonic annihilation operators $\hat{a}_i$. In this approximation, the intracavity photon number and energy in $i$-th mode thus are $N_i(t) = \langle \hat{a}_i^\dagger \hat{a}_i \rangle$, $U_i(t) = \hbar \omega_i N_i(t)$. Losses in $N_i$ arise from intrinsic dissipation described by $\kappa_i^{\text{int}}$ (material absorption, surface scattering, radiation leakage) and an (engineered) out-coupling channel to an output waveguide $\kappa_i^{\text{ext}}$. The total loss rate $\kappa_i$ of the energy of the $i$-th cavity mode $\kappa_i = \kappa_i^{\text{int}} + \kappa_i^{\text{ext}}$. The corresponding loaded (total) quality factor is $Q_i = \frac{\omega_i}{\kappa_i}$, while $Q_i^{\text{int}} = \frac{\omega_i}{\kappa_i^{\text{int}}}, Q_i^{\text{ext}} = \frac{\omega_i}{\kappa_i^{\text{ext}}}$, with $1/Q_i = 1/Q_i^{\text{int}} + 1/Q_i^{\text{ext}}$ [1]. The cavity field amplitude $\langle a_i \rangle$ decays at rate $\kappa_i/2$, consistent with photon number decaying at rate $\kappa_i$.

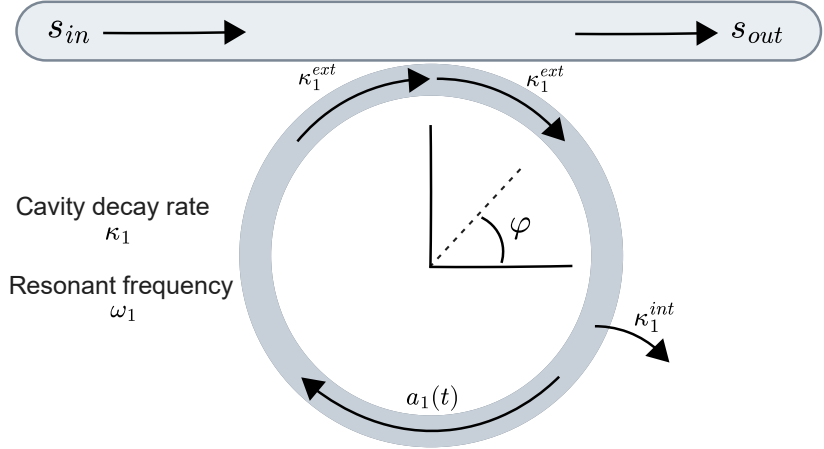

FIG. S1: **Schematic of the system described with coupled-mode theory.**

The aim of this section is to establish a relation between the input photonic flux $\Phi_{\text{in}}^{\text{wg}}$ and the driving term in the Hamiltonian. Following standard temporal coupled mode theory [2], let us consider mode 1 being driven resonantly by an external coherent input from the waveguide (fiber, prism) $s_{\text{in}}(t)$ normalized such that the input photon flux in the waveguide is $\Phi_{\text{in}}^{\text{wg}} \equiv |s_{\text{in}}|^2$ (photons/s), with $\Phi_{\text{in}}^{\text{wg}} = P_{\text{in}}/\hbar \omega_1$ (Fig. S1). Then the amplitude of this mode obeys:

$$\frac{d\langle \hat{a}_1 \rangle}{dt} = -\frac{\kappa_1}{2} \langle \hat{a}_1 \rangle + \sqrt{\kappa_1^{\text{ext}}}\, s_{\text{in}}. \tag{S1}$$

The corresponding input-output relation is $s_{\text{out}} = -s_{\text{in}} + \sqrt{\kappa_1^{\text{ext}}}\, \langle \hat{a}_1 \rangle$.

Then, since $\frac{dN_1}{dt} = \left\langle \frac{d\hat{a}_1^\dagger}{dt} \hat{a}_1 + \hat{a}_1^\dagger \frac{d\hat{a}_1}{dt} \right\rangle$, using Eq. (S1) we get

$$\frac{dN_1}{dt} = -\kappa_1 N_1 + 2\sqrt{\kappa_1^{\text{ext}}}\, \text{Re}[s_{\text{in}}^* \langle \hat{a}_1 \rangle]. \tag{S2}$$

We denote the second term as $\Phi_{\text{in}}^{\text{cav}} \equiv 2\sqrt{\kappa_1^{\text{ext}}}\,\text{Re}[s_{\text{in}}^* \langle \hat{a}_1 \rangle]$ (photons/s). In steady state:

$$\Phi_{\text{in}}^{\text{cav, ss}} = \kappa_1 N_1^{\text{ss}} = \frac{4\kappa_1^{\text{ext}}}{\kappa_1}\Phi_{\text{in}}^{\text{wg,ss}}, \tag{S3}$$

meaning that the total photon loss rate $\kappa_1 N_1$ is balanced by this effective incoming flux $\Phi_{\text{in}}^{\text{cav}}$. This effective flux can be larger than the real waveguide flux $\Phi_{\text{in}}^{\text{wg,ss}}$ necessary to maintain the same intracavity photon number $N_1^{\text{ss}}$, because the destructive interference between two terms in $s_{\text{out}}$ suppresses the loss of the cavity energy via the waveguide.

## S2. OPTONUCLEAR QUADRUPOLAR EFFECT (ONQ) EXCITATION RATE

In this section, we derive an effective second-order optical coupling to the nuclear electric quadrupole moment. The second-order response of the quadrupole tensor to the electric fields $E_p(\omega_{o1}), E_q(\omega_{o2})$ is described by the optonuclear quadrupolar effect (ONQ) [3],

$$H_{\text{ONQ}} = \sum_{ij} \hat{D}_{ij,+}^{pq}\, E_p(\omega_{o1}) E_q(\omega_{o2})\, e^{+i(\omega_{o1} \pm \omega_{o2})t} + \text{H.c.}, \tag{S4}$$

where $\hat{D}_{ij,+}^{pq}$ is the second-order field response in the nuclear quadrupolar interaction tensor ($[D] = \text{J V}^{-2}\,\text{m}^2$), and $E_p(\omega_{o1})(E_q(\omega_{o2}))$ is the $p(q)$-th component of the electric field at frequencies $\omega_{o1}$ ($\omega_{o2}$). Further we will consider the one-color two-photon case $\omega_{o1} = \omega_{o2} = \omega_{\text{iso}}/2$.

From second-order perturbation theory,

$$D = Q_{E2} \sum_{mnl} \frac{e}{4\pi\epsilon_0} \left\langle m \left| \frac{3 r_i r_j - \delta_{ij}\, r^2}{r^5} \right| n \right\rangle \frac{1}{\omega_{mn} - \omega_{\text{iso}}}$$
$$\times 2\left( \frac{f_{lm}\,\langle n|er_p|l\rangle\langle l|er_q|m\rangle}{\omega_{ml} - \omega_p} - \frac{f_{nl}\,\langle n|er_q|l\rangle\langle l|er_p|m\rangle}{\omega_{ln} - \omega_p} \right). \tag{S5}$$

Assuming $\omega_{nl} = \omega_{ml} = \omega_{\text{gap}}$ and approximating the moment operators with $a_0$ (Bohr radius),

$$D = Q_{E2} \frac{2e}{4\pi\epsilon_0} \frac{1}{a_0^3} \frac{1}{\omega_{\text{gap}} - \omega_{\text{iso}}} \frac{e^2 a_0^2}{\omega_{\text{gap}} - \omega_{\text{iso}}/2}, \tag{S6}$$

From Fermi's Golden rule, replacing frequency with energy (and adding $\hbar$), we derive the rate

$$R = \frac{1}{\hbar}\left|\langle g|D\, E_p^2|m\rangle\right|^2 \frac{1}{\hbar 2\Gamma_L} =$$
$$= \frac{1}{\hbar^2}\left|\langle g|Q_{E2}|m\rangle\right|^2 \left( \frac{2e^3}{4\pi\epsilon_0 a_0}\frac{1}{(E_{\text{gap}} - E_{\text{iso}})\left(E_{\text{gap}} - \frac{E_{\text{iso}}}{2}\right)} \right)^2 E_p^4 \frac{1}{2\Gamma_L}, \tag{S7}$$

where $\left|\langle g|Q_{E2}|m\rangle\right|^2 = 30$ W.u. $= 0.376\, e^2\,\text{barn}^2$.

Remembering that $E_p^2 = \frac{2I}{nc\epsilon_0}$ and inserting into the equation above, we get

$$R = \frac{1}{\hbar^2}\left|\langle g|Q_{E2}|m\rangle\right|^2 \left(\frac{2e^3}{4\pi\epsilon_0 a_0}\frac{1}{(E_{\text{gap}} - E_{\text{iso}})\left(E_{\text{gap}} - \frac{E_{\text{iso}}}{2}\right)}\right)^2 \left(\frac{2I}{nc\epsilon_0}\right)^2 \frac{1}{2\Gamma_L}, \quad (S8)$$

Equivalently, in terms of the reduced transition probability $B(E2)$.

$$R = \frac{6\,B(E2)}{\hbar^2 2\Gamma_L}\frac{e^6 I^2}{n^2\pi^2\epsilon_0^4 c^2 a_0^2}\frac{1}{(E_{\text{gap}} - E_{\text{iso}})^2 \left(E_{\text{gap}} - \frac{E_{\text{iso}}}{2}\right)^2}, \quad (S9)$$

where

$$B(E2) = 75\,\epsilon_0 \hbar \left(\frac{\hbar c}{E_{\text{iso}}}\right)^5 \Gamma_{E2}, \quad (S10)$$

Expressing the excitation rate through cross-section, we have $R = \sigma_{\text{ONQ}}\frac{I}{\hbar\omega_{\text{iso}}/2}$. Plugging into the formula above, we derive the cross-section of the ONQ interaction referred to in the main text:

$$\sigma_{\text{ONQ}} = \frac{R\hbar\omega_{\text{iso}}}{2I} = \frac{6\,B(E2)}{\hbar^2 2\Gamma_L}\frac{e^6 I \hbar\omega_{\text{iso}}/2}{n^2\pi^2\epsilon_0^4 c^2 a_0^2}\frac{1}{(E_{\text{gap}} - E_{\text{iso}})^2 \left(E_{\text{gap}} - \frac{E_{\text{iso}}}{2}\right)^2} =$$

$$= \left(\frac{E_{\text{iso}}}{2}\right)\frac{450\epsilon_0 \hbar\left(\frac{\hbar c}{E_{\text{iso}}}\right)^5 \Gamma_{E2}}{\hbar^2\,2\Gamma_L}\frac{e^6 I}{n^2\pi^2\epsilon_0^4 c^2 a_0^2}\frac{1}{(E_{\text{gap}} - E_{\text{iso}})^2\left(E_{\text{gap}} - \frac{E_{\text{iso}}}{2}\right)^2} =$$

$$= \frac{450\hbar^4 c^3 e^6}{2\pi^2 \epsilon_0^3 a_0^2 n^2}\frac{1}{(E_{\text{gap}} - E_{\text{iso}})^2\left(E_{\text{gap}} - \frac{E_{\text{iso}}}{2}\right)^2}\frac{I}{E_{\text{iso}}^4}\frac{\Gamma_{E2}}{2\Gamma_L}. \quad (S11)$$

We can connect the matrix element $\langle g|D|m\rangle$ to $\sigma_{\text{ONQ}}$:

$$\sigma_{\text{ONQ}} = \frac{2\hbar\omega_{\text{iso}}\langle g|D|m\rangle^2}{n^2 c^2 \epsilon_0^2}\frac{I}{2\Gamma_L}. \quad (S12)$$

For CaF$_2$, we calculate $\langle g|D|m\rangle = 1.134 \times 10^{-13}$ Hz V$^{-2}$ m$^2$.

## S3. NUCLEAR ENSEMBLE IN THE CAVITY

### A. Hamiltonian

We consider an ensemble of $N_{\text{Th}}$ $^{229}$Th nuclei embedded in a WGM cavity and interacting with two running-wave modes $\hat{a}_1$ and $\hat{a}_2$ with frequencies $\omega_{1,2}$. In each nucleus we consider only a pair of sublevels $|g\rangle$ and $|e\rangle$ of the ground and isomeric states with transition frequency $\omega_{\text{iso}} \approx \omega_2 \approx 2\omega_1$, and this transition is coupled to the modes $\hat{a}_1$ and $\hat{a}_2$ via two- and single-photon processes, respectively. Transitions involving all other sublevels are supposed to be far from the resonance, and are not considered in this model.

The two-photon coupling between the nucleus and the mode $\hat{a}_1$ is realized via the optonuclear quadrupolar effect (ONQ), whose cross-section is estimated in the previous section, and which can be described by the Hamiltonian (see Eq. (S4)): $\hat{H}_{\text{ONQ}} = \sum_{ij} \hat{D}^{pq}_{ij,+} \hat{E}_p(\omega_1) \hat{E}_q(\omega_1) e^{i(2\omega_1)t} + \text{H.c.}$, where $\hat{E}_p(\omega_1)$ is the $p$-th component of the electric field of the mode $\hat{a}_1$. For simplicity, we neglect the spatial inhomogeneity of the optical modes in the resonator cross-section, as well as diagonal terms in the two-photon interaction between the mode $\hat{a}_1$ and the nuclei (these could lead to a vacuum energy shift). Expressing $\hat{E}_p(\omega_1)$ as a function of $\hat{a}_1$ and omitting diagonal terms, the two-photon interaction between the mode $\hat{a}_1$ and the clock transition in the $j$-th nucleus can be described by the Hamiltonian:

$$\hat{H}^{(0)}_{1,j} = \hbar g \left[ \hat{a}_1 e^{im_1 \varphi_j} + \hat{a}_1^\dagger e^{-im_1 \varphi_j} \right]^2 (\hat{\sigma}^j_{ge} + \hat{\sigma}^j_{eg}).$$

Here, $m_1$ is the azimuthal mode number of the mode $\hat{a}_1$, and the coupling strength $g = \frac{\langle e|\hat{D}|g\rangle E_{\text{zpf}}^2}{\hbar}$, where the value of $\langle e|\hat{D}|g\rangle$ is calculated via Eq. (S6) for CaF$_2$ (similarly, one can calculate this value for other fluoride compounds, such as Na$_2$ThF$_6$ [4]). $E_{\text{zpf}}$ is the electric field corresponding to vacuum in the cavity: $E_{\text{zpf}} = \sqrt{\frac{\hbar \omega_1}{2V_1 \varepsilon_0 n_1^2}}$, where $V_1$ is the mode volume at 296 nm, $n_1$ is the refractive index at the frequency $\omega_1$, and $\varepsilon_0$ is the vacuum permittivity. Electromagnetic mode field distributions and volumes were obtained with finite-difference time-domain simulations for the resonator parameters from the main text.

Similarly, the magnetic dipole coupling between the $j$-th nucleus and the mode $\hat{a}_2$ is described by the Hamiltonian

$$\hat{H}^{(0)}_{2,j} = \hbar G \left[ \hat{a}_2 e^{im_2 \varphi_j} + \hat{a}_2^\dagger e^{-im_2 \varphi_j} \right] (\hat{\sigma}^j_{ge} + \hat{\sigma}^j_{eg}),$$

where $G = \frac{\langle e|\mu|g\rangle B_{\text{zpf}}}{\hbar}$, $B_{\text{zpf}} = \sqrt{\frac{\mu_0 \hbar \omega_2}{2V_2}}$, $V_2$ is the volume of the mode at 148 nm.

Then the interaction Hamiltonian in the Schrödinger picture can be written as:

$$\begin{aligned}\hat{H}^{(0)}_{\text{int}} = &\hbar G \sum_j \left[ \hat{a}_2 e^{im_2 \varphi_j} + \hat{a}_2^\dagger e^{-im_2 \varphi_j} \right] (\hat{\sigma}^j_{ge} + \hat{\sigma}^j_{eg}) \\ &+ \hbar g \sum_j \left[ \hat{a}_1 e^{im_1 \varphi_j} + \hat{a}_1^\dagger e^{-im_1 \varphi_j} \right]^2 (\hat{\sigma}^j_{ge} + \hat{\sigma}^j_{eg}),\end{aligned} \quad (S13)$$

where $G$ and $g$ are single- and two-photon coupling strengths, $x_j$ is the position of the $j$-th nucleus along the equator of the cavity, $k_{1,2}$ are the effective wave numbers of the modes $\hat{a}_1$ and $\hat{a}_2$, $\hat{\sigma}^j_{\alpha\beta} = |\alpha_j\rangle\langle\beta_j| \bigotimes_{l \neq j} \mathbb{1}_l$ is the transition operator between states $\alpha$ and $\beta$ of the $j$-th nucleus, $|e\rangle$, $|g\rangle$ are the isomer and ground states of the nucleus, respectively.

The full Hamiltonian $\hat{H}^{(0)}$ of the system in the Schrödinger picture is

$$\begin{aligned}\hat{H}^{(0)} = &\hbar\left[\omega_1 \hat{a}_1^\dagger \hat{a}_1 + \omega_2 \hat{a}_2^\dagger \hat{a}_2 + \omega_{\text{iso}}\sum_j \hat{\sigma}_{ee}^j\right] + \hat{H}_{\text{int}}^{(0)} \\ &+ \hbar\left[\mathcal{E}\hat{a}_1^\dagger e^{-i(\omega_L t + \phi(t))} + \mathcal{E}^* \hat{a}_1 e^{i(\omega_L t + \phi(t))}\right],\end{aligned} \tag{S14}$$

where the last term corresponds to a coherent drive of the external field with frequency $\omega_L$ and a fluctuating phase $\phi(t)$, which can be considered as a Wiener process with $\langle\dot\phi(t)\dot\phi(t')\rangle = \Gamma_L \delta(t-t')$, and $\Gamma_L$ is the linewidth of the driving laser (FWHM).

Performing unitary transformations, we can further simplify the Hamiltonian:

$$\hat{H} = i\hbar \frac{\partial \hat{U}^\dagger}{\partial t}\hat{U} + \hat{U}^\dagger \hat{H}^{(0)} \hat{U}. \tag{S15}$$

First, we apply the unitary matrix

$$\hat{U}_1 = \exp\left[-i\omega_L\left(\hat{a}_1^\dagger \hat{a}_1 + 2\hat{a}_2^\dagger \hat{a}_2 + 2\sum_j \hat{\sigma}_{ee}^j\right)t\right]. \tag{S16}$$

Omitting fast oscillating terms, we obtain

$$\begin{aligned}\hat{H}^{(1)} =& \hbar\left[\delta_1 \hat{a}_1^\dagger \hat{a}_1 + \delta_2 \hat{a}_2^\dagger \hat{a}_2 + \Delta \sum_j \hat{\sigma}_{ee}^j\right] + \hbar\left[\mathcal{E}\hat{a}_1^\dagger e^{-i\phi(t)} + \mathcal{E}^* \hat{a}_1 e^{i\phi(t)}\right] \\ &+ \hbar G \sum_j\left[\hat{\sigma}_{eg}^j \hat{a}_2 e^{im_2\varphi_j} + \hat{\sigma}_{ge}^j \hat{a}_2^\dagger e^{-im_2\varphi_j}\right] + \hbar g \sum_j\left[\hat{\sigma}_{eg}^j \hat{a}_1^2 e^{2im_1\varphi_j} + \hat{\sigma}_{ge}^j \hat{a}_1^{\dagger 2} e^{-2im_1\varphi_j}\right],\end{aligned} \tag{S17}$$

where $\delta_1 = \omega_1 - \omega_L$, $\delta_2 = \omega_2 - 2\omega_L$, $\Delta = \omega_{\text{iso}} - 2\omega_L$.

Now we perform one more unitary transformation Eq. (S15) with the unitary matrix

$$\hat{U}_2 = \exp\left[2im_1\sum_j \hat{\sigma}_{ee}^j \varphi_j\right], \tag{S18}$$

which gives us

$$\begin{aligned}\hat{H}^{(2)} =& \hbar\left[\delta_1 \hat{a}_1^\dagger \hat{a}_1 + \delta_2 \hat{a}_2^\dagger \hat{a}_2 + \Delta \sum_j \hat{\sigma}_{ee}^j\right] + \hbar\left[\mathcal{E}\hat{a}_1^\dagger e^{-i\phi(t)} + \mathcal{E}^* \hat{a}_1 e^{i\phi(t)}\right] \\ &+ \hbar G \sum_j\left[\hat{\sigma}_{eg}^j \hat{a}_2 e^{\psi_j} + \hat{\sigma}_{ge}^j \hat{a}_2^\dagger e^{-i\psi_j}\right] + \hbar g \sum_j\left[\hat{\sigma}_{eg}^j \hat{a}_1^2 + \hat{\sigma}_{ge}^j \hat{a}_1^{\dagger 2}\right].\end{aligned} \tag{S19}$$

Here $\psi_j = (m_2 - 2m_1)\varphi_j \ne 0$, due to the difference in effective refractive indices $n_1^{\text{eff}}$ and $n_2^{\text{eff}}$ of the WGM modes, where neglecting the geometric dispersion one has $m_i \approx \frac{\omega_i n_i R}{c}$[5].

## B. Relaxations

In the Heisenberg representation, an operator $\hat{o}$ follows the Heisenberg-Langevin equation

$$\frac{d\hat{o}}{dt} = \frac{i}{\hbar}[\hat{H}, \hat{o}] + \hat{\mathcal{L}}[\hat{o}] + \hat{\mathcal{F}}(\hat{o}), \tag{S20}$$

where $\hat{\mathcal{F}}_{\hat{o}}$ is a Langevin force, and the Liouvillian term $\hat{\mathcal{L}}[\hat{o}]$ describing the dissipative process can be written as

$$\hat{\mathcal{L}}[\hat{o}] = \sum_k \gamma_k \hat{\mathcal{D}}[\hat{J}_k]\hat{o}, \quad \text{where} \quad \hat{\mathcal{D}}[\hat{J}]\hat{o} = \hat{J}^\dagger \hat{o} \hat{J} - \frac{1}{2}\left(\hat{J}^\dagger \hat{J}\hat{o} + \hat{o}\hat{J}^\dagger \hat{J}\right). \tag{S21}$$

Here $\hat{\mathcal{D}}$ is a *Lindbladian superoperator*, $\hat{J}_k$ are *jump operators* and $\gamma_k$ are *relaxation rates*. In our model the relaxation processes are described by the Liouvillian

$$\hat{\mathcal{L}}^{(2)}[\hat{o}] = \kappa_1 \hat{\mathcal{D}}[\hat{a}_1]\hat{o} + \kappa_2 \hat{\mathcal{D}}[\hat{a}_2]\hat{o} + \sum_j \left[\gamma \hat{\mathcal{D}}[\hat{\sigma}_{ge}^j]\hat{o} + \nu \hat{\mathcal{D}}[\hat{\sigma}_{ee}^j]\hat{o}\right], \tag{S22}$$

where $\gamma = \Gamma_{\mathrm{M1}} n_2^3$ is the spontaneous decay rate of the isomeric state in the crystal lattice environment, $\nu$ is the dephasing rate (fundamentally limited in $\mathrm{CaF}_2$ by the interaction with magnetic momenta of nearby fluorine nuclei) [6], $\kappa_{1,2}$ are the total energy decay rates of the modes $\hat{a}_1$ and $\hat{a}_2$ respectively ($\kappa_{1,2} = \kappa_{1,2}^{\mathrm{ext}} + \kappa_{1,2}^{\mathrm{int}}$).

The Hamiltonian Eq. (S19) contains a fluctuating phase $\phi(t)$. Following the approach used in Refs. [7, 8], we perform a transformation Eq. (S15) with a unitary matrix

$$\hat{U} = \exp\left[-i\left(\hat{a}_1^\dagger \hat{a}_1 + 2\hat{a}_2^\dagger \hat{a}_2 + 2\sum_j \hat{\sigma}_{ee}^j\right)\phi(t)\right] \tag{S23}$$

to the Hamiltonian Eq. (S19). The new Hamiltonian $\hat{H}$ has the form $\hat{H} = \hat{H}^D + \hat{H}^S \dot{\phi}$, where the deterministic part $\hat{H}^D$ is equal to

$$\begin{aligned}\hat{H}^D = &\hbar\left[\delta_1 \hat{a}_1^\dagger \hat{a}_1 + \delta_2 \hat{a}_2^\dagger \hat{a}_2 + \Delta \sum_j \hat{\sigma}_{ee}^j\right] + \hbar\left[\mathcal{E}\hat{a}_1^\dagger + \mathcal{E}^*\hat{a}_1\right] \\ &+ \hbar G \sum_j \left[\hat{\sigma}_{eg}^j \hat{a}_2 e^{i\psi_j} + \hat{\sigma}_{ge}^j \hat{a}_2^\dagger e^{-i\psi_j}\right] + \hbar g \sum_j \left[\hat{\sigma}_{eg}^j \hat{a}_1^2 + \hat{\sigma}_{ge}^j \hat{a}_1^{\dagger 2}\right],\end{aligned} \tag{S24}$$

and the stochastic part is

$$\hat{H}^S = -\hbar\left[\hat{a}_1^\dagger \hat{a}_1 + 2\hat{a}_2^\dagger \hat{a}_2 + 2\sum_j \hat{\sigma}_{ee}^j\right]. \tag{S25}$$

Now, the Heisenberg-Langevin equation Eq. (S20) can be written in the form

$$(S) \; d\hat{o} = \frac{i}{\hbar}[\hat{H}^D, \hat{o}]dt + \frac{i}{\hbar}[\hat{H}^S, \hat{o}]\sqrt{\Gamma_L}\, dW + \hat{\mathcal{L}}^{(2)}[\hat{o}]dt + \hat{\mathcal{F}}(\hat{o})dt, \quad (S26)$$

which needs to be interpreted as a Stratonovic stochastic differential equation (indicated by (S)). Here $W$ is a $\delta$-correlated Wiener process, $\langle dW(t)dW(t')\rangle = \delta(t-t')$. This equation can be transformed into Ito form (indicated by (I)) as

$$(I) \; d\hat{o} = \frac{i}{\hbar}[\hat{H}^D, \hat{o}]dt + \frac{i}{\hbar}[\hat{H}^S, \hat{o}]\sqrt{\Gamma_L}\, dw + \frac{\Gamma_L}{\hbar^2}\hat{\mathfrak{D}}[\hat{H}^S]\hat{o}\, dt + \hat{\mathcal{L}}^{(2)}[\hat{o}]dt + \hat{\mathcal{F}}(\hat{o})dt. \quad (S27)$$

Ensemble averaging of this equation leads to the disappearance of both stochastic parts (the Langevin force and the phase fluctuations), and we conclude that the averaged value of some system operator $\hat{o}$ follows the equation

$$\frac{d\langle \hat{o}\rangle}{dt} = \frac{i}{\hbar}\langle[\hat{H}, \hat{o}]\rangle + \langle \hat{\mathcal{L}}[\hat{o}]\rangle, \quad (S28)$$

with Hamiltonian $\hat{H} = \hat{H}^D$ (see Eq. (S24)) and Lindbladian

$$\hat{\mathcal{L}}[\hat{o}] = \kappa_1 \hat{\mathfrak{D}}[\hat{a}_1]\hat{o} + \kappa_2 \hat{\mathfrak{D}}[\hat{a}_2]\hat{o} + \sum_j \left[\gamma \hat{\mathfrak{D}}[\hat{\sigma}_{ge}^j]\hat{o} + \nu \hat{\mathfrak{D}}[\hat{\sigma}_{ee}^j]\hat{o}\right]$$
$$+ \Gamma_L \hat{\mathfrak{D}}\left[\hat{a}_1^\dagger \hat{a}_1 + 2\hat{a}_2^\dagger \hat{a}_2 + 2\sum_j \hat{\sigma}_{ee}^j\right]\hat{o}. \quad (S29)$$

### C. Cumulant expansion

Let us start from the equations for the averaged values of the most relevant system operators. Without loss of generality, we take $\mathcal{E} = \mathcal{E}^* = \sqrt{\kappa_1^{\text{ext}}} s_{\text{in}} = \sqrt{\kappa_1^{\text{ext}} \Phi_{\text{in}}^{\text{wg}}}$, which corresponds to the choice of $e^{-i\alpha} = -i$ in Eq. (S1). The cavity mode $\langle \hat{a}_1 \rangle$ follows the equation

$$\frac{d\langle \hat{a}_1\rangle}{dt} = -i\mathcal{E} - \left(\frac{\kappa_1 + \Gamma_L}{2} + i\delta_1\right)\langle \hat{a}_1\rangle - 2ig\sum_j \langle \hat{a}_1^\dagger \hat{\sigma}_{ge}^j\rangle. \quad (S30)$$

In this expression the first term $-i\mathcal{E}$ is much larger than the last one. Indeed, taking the quality factor $Q_1 = 5 \times 10^6$ (see Table S1), supposing that $\kappa_1^{\text{ext}} = \kappa_1/2 \approx 2\pi \times 101$ MHz, and taking a rather small input power $P_{\text{in}} = 100\,\mu$W of 296 nm light delivered to the cavity via the waveguide, we obtain $\mathcal{E} \approx 3.08 \times 10^{11}\,\text{s}^{-1}$, whereas with the typical value of $g = 5.14 \times 10^{-8}\,\text{s}^{-1}$ and $N_{\text{Th}}^1 = V_1 \rho = 4.66 \times 10^{10}$ we find that the last term can not exceed $1.7 \times 10^3\,\text{s}^{-1}$. Also, experimentally feasible values of detuning $\delta_1$ and linewidth $\Gamma_L$ of the pumping laser are much smaller than $\kappa_1$,

therefore, the mode $\hat{a}_1$ will be defined by the driving field $\mathcal{E}$ and can be approximated by a classical $c$-number coherent field $a_1$.

Similarly, equations for the second cavity mode $\langle \hat{a}_2 \rangle$ and individual polarizations $\langle \hat{\sigma}_{ge}^j \rangle$ are

$$\frac{d\langle \hat{a}_2 \rangle}{dt} = -\left(\frac{\kappa_2}{2} + 2\Gamma_L + i\delta_2\right)\langle \hat{a}_2 \rangle - iG\sum_j \langle \hat{\sigma}_{ge}^j \rangle e^{-i\psi_j}, \tag{S31}$$

$$\frac{d\langle \hat{\sigma}_{ge}^j \rangle}{dt} = -\left[i\Delta + \frac{\gamma + \nu}{2} + 2\Gamma_L\right]\langle \hat{\sigma}_{ge}^j \rangle + iga_1^2 \langle \hat{\sigma}_z^j \rangle + iG \langle \hat{\sigma}_z^j \hat{a}_2 e^{i\psi_j} \rangle, \tag{S32}$$

We suppose that the distribution of phases $\psi_j$ is uniform in $[0, 2\pi)$. Then the contributions from polarizations $\langle \hat{\sigma}_{eg}^j \rangle$ of different dipoles will compensate each other, and the field of the second mode is $\langle \hat{a}_2 \rangle \approx 0$. This field is not coherent, and $N_2 = \langle \hat{a}_2^\dagger \hat{a}_2 \rangle \neq |\langle \hat{a}_2 \rangle|^2$. We can write the equations

$$\frac{dN_2}{dt} = -\kappa_2 N_2 + iG\sum_j \left(\langle \hat{\sigma}_{eg}^j \hat{a}_2 \rangle e^{i\psi_j} - \langle \hat{\sigma}_{ge}^j \hat{a}_2^\dagger \rangle e^{-i\psi_j}\right), \tag{S33}$$

$$\begin{aligned}\frac{d\langle \hat{a}_2^\dagger \hat{\sigma}_{ge}^j \rangle}{dt} = & -\left[i(\Delta - \delta_2) + \frac{\kappa_2 + \gamma + \nu}{2}\right]\langle \hat{a}_2^\dagger \hat{\sigma}_{ge}^j \rangle + iga_1^2 \langle \hat{a}_2^\dagger \hat{\sigma}_z^j \rangle \\ & + iG\left\{\sum_{k\neq j}\langle \hat{\sigma}_{eg}^k \hat{\sigma}_{ge}^j \rangle e^{i\psi_k} + e^{i\psi_j}\left(\langle \hat{\sigma}_z^j \hat{a}_2^\dagger \hat{a}_2 \rangle + \frac{1 + \langle \hat{\sigma}_z^j \rangle}{2}\right)\right\},\end{aligned} \tag{S34}$$

where $\hat{\sigma}_z^j = \hat{\sigma}_{ee}^j - \hat{\sigma}_{gg}^j$ is the inversion, and $\langle \hat{\sigma}_{eg}^k \hat{\sigma}_{ge}^j \rangle$ is the inter-nuclear coherence, and

$$\begin{aligned}\frac{d\langle \hat{\sigma}_z^j \rangle}{dt} = & -\gamma(\langle \hat{\sigma}_z^j \rangle + 1) + 2ig\left(a_1^{*2}\langle \hat{\sigma}_{ge}^j \rangle - a_1^2 \langle \hat{\sigma}_{eg}^j \rangle\right) \\ & + 2iG\left(\langle \hat{a}_2^\dagger \hat{\sigma}_{ge}^j \rangle e^{-i\psi_j} - \langle \hat{a}_2 \hat{\sigma}_{eg}^j \rangle e^{i\psi_j}\right),\end{aligned} \tag{S35}$$

$$\begin{aligned}\frac{d\langle \hat{\sigma}_{eg}^k \hat{\sigma}_{ge}^j \rangle}{dt} = & -(\gamma + \nu)\langle \hat{\sigma}_{eg}^k \hat{\sigma}_{ge}^j \rangle + ig\left(a_1^2 \langle \hat{\sigma}_z^j \hat{\sigma}_{eg}^k \rangle - a_1^{*2}\langle \hat{\sigma}_z^k \hat{\sigma}_{ge}^j \rangle\right) \\ & + iG\left(e^{i\psi_j}\langle \hat{\sigma}_z^j \hat{\sigma}_{eg}^k \hat{a}_2 \rangle - e^{-i\psi_k}\langle \hat{\sigma}_z^k \hat{\sigma}_{eg}^j \hat{a}_2^\dagger \rangle\right).\end{aligned} \tag{S36}$$

Finally, the number of photons $N_1 = \langle \hat{a}_1^\dagger \hat{a}_1 \rangle$ in the cavity mode $a_1$ follows the equation

$$\frac{dN_1}{dt} = -\kappa_1 N_1 + i\mathcal{E}(a_1 - a_1^*) + 2ig\sum_j \left(a_1^2 \langle \hat{\sigma}_{eg}^j \rangle - a_1^{*2}\langle \hat{\sigma}_{ge}^j \rangle\right). \tag{S37}$$

Equations Eq. (S30) – Eq. (S37) do not form a closed set, because they contain third-order correlators, and the equations for the latter would contain 4th-order correlators, etc., which makes this system computationally intractable without further approximations. One possible method is to use a cumulant approximation, where cumulants above a certain order are neglected, and

higher-order correlators are expressed via lower-order ones [9]. We have performed numerical simulation with the help of Julia package QuantumCumulants.jl [10], using second-order cumulant approximation and grouping the nuclear ensemble into $M = 10$ clusters with different phases $\phi_k, k = 0, 1, \ldots, M$ (same for all nuclei in one cluster). These calculations confirmed the results obtained with the help of more simple semiclassical equations which will be presented below (see section S3 E).

### D. Rate equations

First we perform the adiabatic elimination of $a_1$, coherences $\langle \hat{\sigma}_{ge}^j \rangle$ and correlations $\langle \hat{a}_2^\dagger \hat{\sigma}_{ge}^j \rangle$.

$$a_1 = \frac{-2i\mathcal{E}}{\kappa_1 + \Gamma_L + 2i\delta_1}, \tag{S38}$$

$$\langle \hat{\sigma}_{ge}^j \rangle = \frac{2iga_1^2 \langle \hat{\sigma}_z^j \rangle}{4\Gamma_L + \gamma + \nu + 2i\Delta}, \tag{S39}$$

$$\langle \hat{a}_2^\dagger \hat{\sigma}_{ge}^j \rangle = \frac{2iGe^{i\psi_j} \left( N_2 \langle \hat{\sigma}_z^j \rangle + \frac{1+\langle \hat{\sigma}_z^j \rangle}{2} \right)}{\kappa_2 + \gamma + \nu + 2i(\Delta - \delta_2)}. \tag{S40}$$

To obtain these expressions, we neglected correlations between the field $\hat{a}_2$ and inversions $\hat{\sigma}_z^j$ in Eqs. (S32) and (S36), as well as the sum over inter-nuclear correlations $\langle \hat{\sigma}_{eg}^k \hat{\sigma}_{ge}^j \rangle$ in Eq. (S34), because the terms with different phases compensate each other. Also, we replaced $\langle \hat{a}_2^\dagger \hat{a}_2 \hat{\sigma}_z^j \rangle$ by $N_2 \langle \hat{\sigma}_z^j \rangle$. Substituting Eq. (S38) – Eq. (S40) into Eqs. (S33), (S35) and (S37), and introducing $S_z = \sum_j \langle \hat{\sigma}_z^j \rangle$, where the sum is taken over $N_{\text{Th}} = \rho V_2$ nuclei overlapping with both cavity modes, we obtain the rate equations:

$$\frac{dN_1}{dt} = -\kappa_1 N_1 + \frac{4|\mathcal{E}|^2 (\kappa_1 + \Gamma_L)}{(\kappa_1 + \Gamma_L)^2 + 4\delta_1^2} + \frac{8g^2 N_1^2 S_z (\gamma + \nu + 4\Gamma_L)}{(\gamma + \nu + 4\Gamma_L)^2 + 4\Delta^2}, \tag{S41}$$

$$\frac{dN_2}{dt} = -\kappa_2 N_2 + \frac{2G^2 (\kappa_2 + \gamma + \nu)[S_z(2N_2 + 1) + N_{\text{Th}}]}{(\kappa_2 + \gamma + \nu)^2 + 4(\Delta - \delta_2)^2}, \tag{S42}$$

$$\frac{dS_z}{dt} = -\gamma(S_z + N_{\text{Th}}) - \frac{8g^2 N_1^2 S_z (\gamma + \nu + 4\Gamma_L)}{(\gamma + \nu + 4\Gamma_L)^2 + 4\Delta^2} - \frac{4G^2 [S_z(2N_2 + 1) + N_{\text{Th}}](\kappa_2 + \gamma + \nu)}{(\kappa_2 + \gamma + \nu)^2 + 4(\Delta - \delta_2)^2}. \tag{S43}$$

Let us introduce the following notations:

$$2\Gamma_1 = \gamma + \nu + 4\Gamma_L + \frac{4\Delta^2}{\gamma + \nu + 4\Gamma_L}, \qquad \Phi_{\text{in}}^{\text{cav}} = \frac{4\mathcal{E}^2(\kappa_1 + \Gamma_L)}{(\kappa_1 + \Gamma_L)^2 + 4\delta_1^2},$$

$$\Gamma_2 = \kappa_2 + \gamma + \nu + \frac{4(\Delta - \delta_2)^2}{\kappa_2 + \gamma + \nu}, \tag{S44}$$

so we can rewrite the equations Eq. (S41) – Eq. (S43) in the form

$$\frac{dN_1}{dt} = -\kappa_1 N_1 + \Phi_{\text{in}}^{\text{cav}} + \frac{4g^2}{\Gamma_1} N_1^2 S_z, \tag{S45}$$

$$\frac{dN_2}{dt} = -\kappa_2 N_2 + \frac{4G^2}{\Gamma_2} N_2 S_z + \frac{2G^2}{\Gamma_2}(N_{\text{Th}} + S_z), \tag{S46}$$

$$\frac{dS_z}{dt} = -\left(\gamma + \frac{4G^2}{\Gamma_2}\right)(S_z + N_{\text{Th}}) - \frac{4g^2}{\Gamma_1} N_1^2 S_z - \frac{8G^2}{\Gamma_2} N_2 S_z. \tag{S47}$$

The generalization of these expressions to the case of inhomogeneous couplings $g$ and $G$ is straightforward if the phase distribution is not correlated with the distribution of magnitudes of $g$ and $G$. In other words, it allows us to divide the nuclear ensemble into $N_C$ clusters such that in each cluster $C_i$ the coefficients $g$ and $G$ can be approximated by their mean values $g_i$ and $G_i$, but summation over phases will again lead to the disappearance of the sum over inter-nuclear correlations $\sum_{k \in C_i, j \in C_m} \langle \hat{\sigma}_{eg}^k \hat{\sigma}_{ge}^j \rangle$, both for the same cluster ($C_i = C_m$) or for two different clusters. In the present paper we consider only such systems, thus establishing the basis for equations (5) in the main text.

Let us verify that the $\Phi_{\text{in}}^{\text{cav}}$ from Eq. (S44) coincides with the previously defined $\Phi_{\text{in}}^{\text{cav}}$ in Eq. (S2). Comparing Eq. (S2) to Eq. (S41) and Eq. (S45), supposing the resonant case ($\delta_1 \ll \kappa_1$) and $\Gamma_L \ll \kappa_1$, we see that since $-i\mathcal{E} \equiv \sqrt{\kappa_1^{\text{ext}}} s_{\text{in}} \implies \mathcal{E} = i\sqrt{\kappa_1^{\text{ext}}} s_{\text{in}}$, $\mathcal{E}^* = -i\sqrt{\kappa_1^{\text{ext}}} s_{\text{in}}^*$, we get in the steady-state (ignoring nuclear back-action on the mode $a_1$):

$$\frac{4|\mathcal{E}|^2}{\kappa_1} = i\mathcal{E}^* a_1 - i\mathcal{E} a_1^* = i(-i\sqrt{\kappa_1^{\text{ext}}} s_{\text{in}}^*) a_1 - i(i\sqrt{\kappa_1^{\text{ext}}} s_{\text{in}}) a_1^* = \tag{S48}$$

$$= \sqrt{\kappa_1^{\text{ext}}} s_{\text{in}}^* a_1 + \sqrt{\kappa_1^{\text{ext}}} s_{\text{in}} a_1^* = 2\sqrt{\kappa_1^{\text{ext}}} \text{Re}[s_{\text{in}}^* a_1] = \Phi_{\text{in}}^{\text{cav}}. \tag{S49}$$

So, one sees that $\Phi_{\text{in}}^{\text{cav}} = \frac{4\kappa_1^{\text{ext}}}{\kappa_1} \Phi_{\text{in}}^{\text{wg}}$.

At critical coupling, $\kappa_1^{\text{ext}} = \kappa_1^{\text{int}} = \kappa_1/2$, this reduces to

$$\Phi_{\text{in}}^{\text{cav}} = 2\Phi_{\text{in}}^{\text{wg}}, \tag{S50}$$

thus, providing us with the link between the experimentally controlled input power in the waveguide and the driving terms appearing in the equations of motion. For simulations presented in this work, critical coupling for both modes is assumed. The output photon flux into the waveguide from the cavity mode is calculated as $\Phi_2^{\text{out}} = \kappa_2^{\text{ext}} N_2 = \kappa_2 N_2/2$ and is thus incoherent created by fluorescence of thorium nuclei inside the cavity. The isotropic emission rate is calculated as $\gamma N_e = \gamma(S_z + N_{\text{Th}})/2$. Note that generally, coherent buildups are destroyed by the phase mismatch between WGM modes at two frequencies.

The table with the parameters used in the main text is presented below.

TABLE S1: Numerical parameters

| $\rho$, cm$^{-3}$ | $Q_1$ | $Q_2$ | $\gamma/2\pi$, kHz | $\gamma_z/2\pi$, kHz | $\Gamma_L/2\pi$, kHz | $V_1$, m$^3$ | $V_2$, m$^3$ |
|---|---|---|---|---|---|---|---|
| $2\times 10^{17}$ | $5\times 10^6$ | $5\times 10^6/8$ | $2.54\times 10^{-7}$ | 0.3 | 1 | $2.33\times 10^{-13}$ | $0.97\times 10^{-13}$ |

| $\langle D \rangle$, Hz V$^{-2}$ m$^2$ | $G/2\pi$, Hz | $g/2\pi$, Hz | $P_{\text{in}}$, mW |
|---|---|---|---|
| $1.134\times 10^{-13}$ | 3.29 | $8.18\times 10^{-9}$ | 100 |

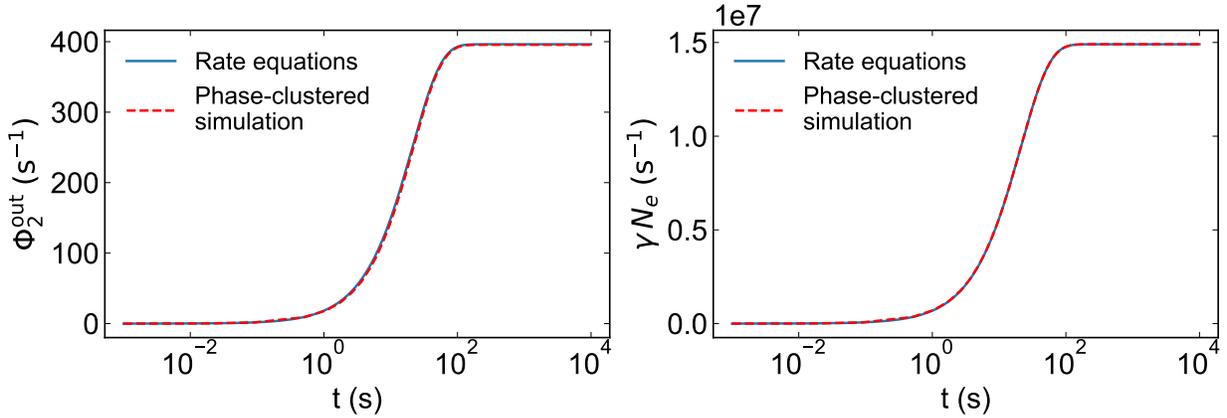

FIG. S2: **Numerical comparison of simulation methods.** Comparison between emission dynamics simulated with rate equations and phase-clustered second order cumulant expansion simulation with the help of QuantumCumulants.jl package.

### E. Clustering approach simulations with Julia package

Below, we compare the results for the cavity output photon flux and isotropic emission rates obtained from rate equations (shown in Fig. 2 of the main text) and corresponding full second-order cumulant expansion results, where we group the nuclear ensemble into $M = 10$ clusters, and inside each cluster all nuclei have a phase $\psi_k$, $k = 0, 1, \ldots, M$ (as defined in Eq. (S19)), uniformly distributed in the range $[0, 2\pi)$ across the clusters. A great match between the approaches is achieved, corroborating the validity of assumptions underlying the derivation of rate equations.

### F. Parameter sweeps

In Fig. S3, we present the dependences of cavity output photon flux $\Phi_2^{\text{out}}$ on the pumping laser linewidth $\Gamma_L$, $\rho$, and $Q_2$ ($Q_1$ changed as well according to $Q_1 = 8Q_2$ based on the scaling $Q \propto \lambda^3 \sigma^{-2}$ with wavelength $\lambda$). Those simulations correspond to the scenario of perfect mode

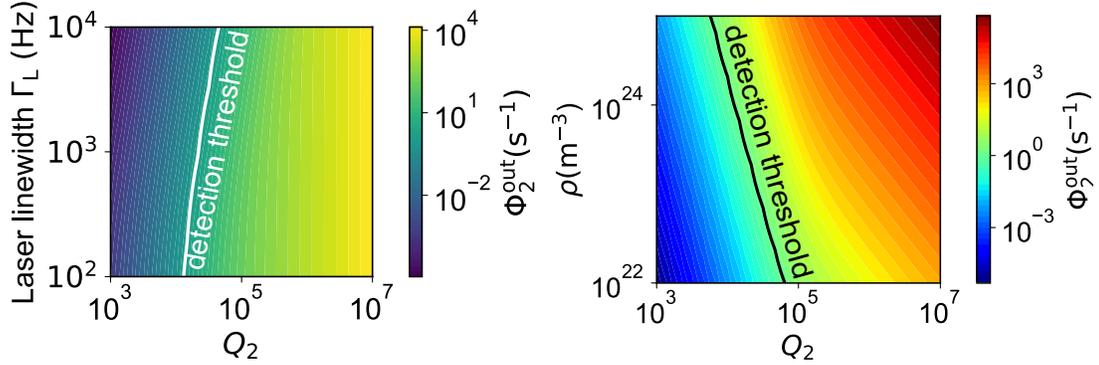

FIG. S3: **Simulated steady-state cavity output photon flux of the 148 nm mode with a pump power of 100 mW**. (a) Dependence of $\Phi_2^{\text{out}}$ on the pumping laser linewidth $\Gamma_L$ ($10^2 - 10^4$ Hz) and the quality factor of the 148 nm mode $Q_2$ ($10^3 - 10^7$). (b) Dependence of $\Phi_2^{\text{out}}$ on the $^{229}$Th nuclei density $\rho$ ($10^{16} - 10^{19}$ cm$^{-3}$) and $Q_2$ ($10^3 - 10^7$).

overlap and uniform $^{229}$Th doping, with total number of interacting nuclei being $N_{\text{Th}} = \rho V_2$.

Fig. S4 shows output photon fluxes for the cases of doping, 100 keV channeled implantation, and 20 keV random implantation, with numerically simulated electromagnetic field distributions. The dashed red line represents the detection threshold, assumed here to be 10 photons/s for isotropic emission and 1 photons/s for outcoupling from the resonator.

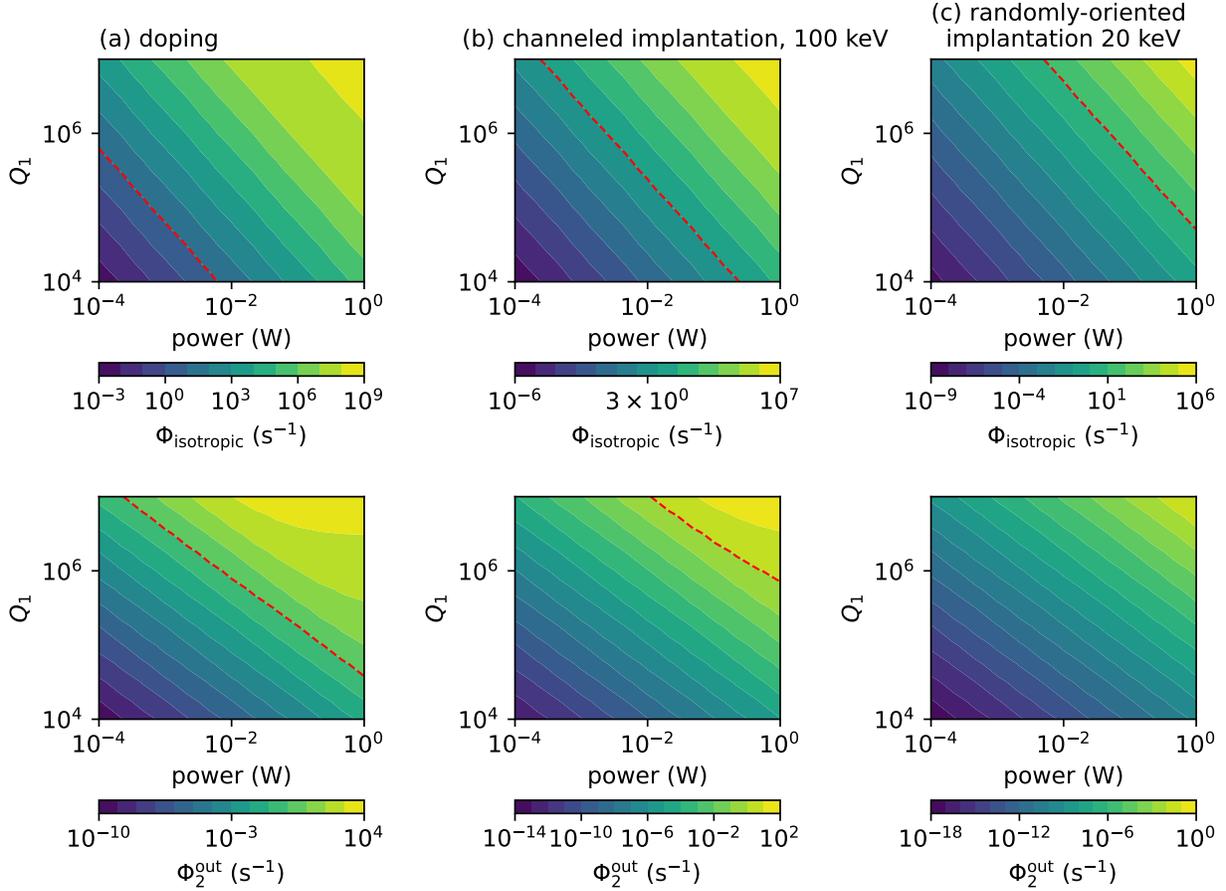

FIG. S4: **Simulated steady-state cavity output photon flux of the** 148 **nm mode and isotropic emission rate as a function of cavity** $Q$**-factor of the** 296 **nm mode and pump power.** (a) Uniform doping. (b) 100 keV channeled implantation. (c) 20 keV randomly-oriented implantation. Top row corresponds to isotropic emission rate; bottom row to output photon flux in waveguide.

### S4. SIMULATION OF RANDOM AND CHANNELED ION IMPLANTATION

Fig. S5 shows implantation depth distributions of $^{229}$Th ions simulated using MARLOWE, for three acceleration energies (20 keV, 100 keV, and 1 MeV), using a Frenkel pair recombination radius of 5.4 Å, comparing random and channeled incidence. Increasing ion energy leads to progressively deeper implantation profiles, while channeling results in a broader and deeper penetration relative to the random case. For reference, the electric field strength at the vertical center of the optical mode ($y = 0$) is overlaid to indicate the spatial overlap between the implanted ions and the guided mode.

The rate equations, obtained in Section S3 D, can be generalized to the case of inhomoge-

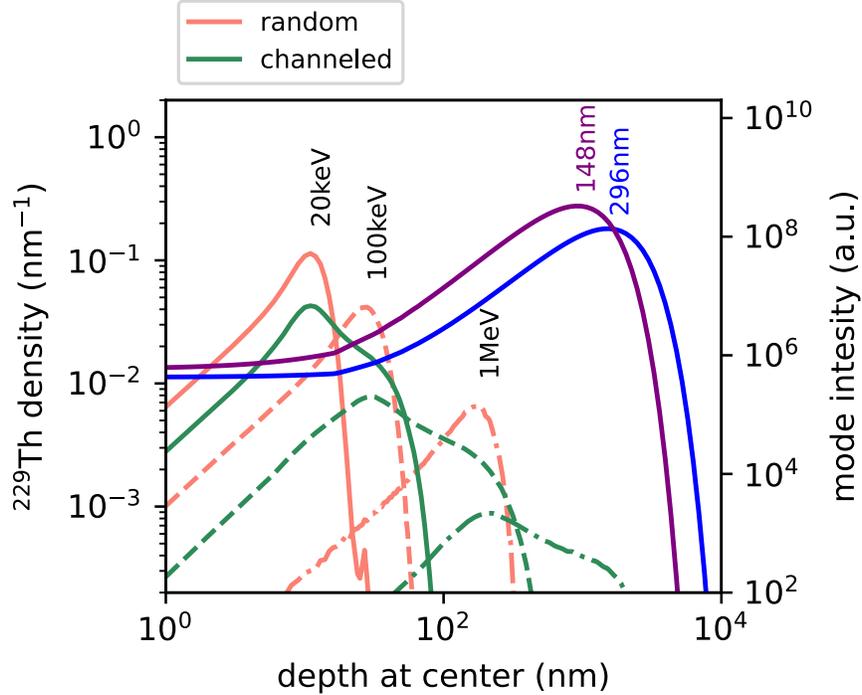

FIG. S5: **Implantation depth distribution for $^{229}$Th ion beams.** Acceleration energies of 20 keV, 100 keV and 1 MeV in random and channeled configurations are shown together with the electric field strength at the vertical center of the mode ($y = 0$).

neous nuclear density distribution $\rho(\vec{r})$ and electromagnetic field profiles. For simplicity, let us assume the pumping field almost immediately reaches steady state and stays approximately constant (which is true to a high precision), and is connected with the photon number $N_1$ as $E_1^2 = 4E_{\text{zpf}}^2 N_1^{ss} = \frac{2\hbar\omega_1}{V_1 \varepsilon_0 n_1^2} N_1^{ss}$, so that $4g^2 N_1^{ss} = D^2 E_1^4/4$ ($\langle e|\hat{D}|g\rangle =: D$). Then we neglect the stimulated emission at 148 nm (because of a small number of intracavity photons of that mode). Thus, in the steady state, we obtain:

$$N_2 = \int \frac{2G^2(\vec{r})}{\kappa_2 \Gamma_2} \frac{\frac{D^2 E_1^4(\vec{r})}{4\Gamma_1}}{\left(\gamma + \frac{D^2 E_1^4(\vec{r})}{4\Gamma_1}\right)} \rho(\vec{r}) dV, \quad (S51)$$

which depends on the spatial overlap between the two modes and the nuclear distribution.

## References

[1] S. Fujii, Y. Hayama, K. Imamura, H. Kumazaki, Y. Kakinuma, and T. Tanabe, Optica **7**, 694 (2020).

[2] J. Joannopoulos, S. Johnson, J. Winn, and R. Meade, *Photonic Crystals: Molding the Flow of Light - Second Edition* (Princeton University Press, 2011).

[3] H. Xu, C. Li, G. Wang, H. Wang, H. Tang, A. R. Barr, P. Cappellaro, and J. Li, Physical Review X **13**, 011017 (2023).

[4] H. Xu, H. Tang, G. Wang, C. Li, B. Li, P. Cappellaro, and J. Li, Physical Review A **108**, L021502 (2023).

[5] I. Breunig, B. Sturman, F. Sedlmeir, H. G. L. Schwefel, K. Buse, M. Spillane, D. K. Armani, and K. J. Vahala, Optics Express **21**, 30683 (2013).

[6] G. A. Kazakov, A. N. Litvinov, V. I. Romanenko, L. P. Yatsenko, A. V. Romanenko, M. Schreitl, G. Winkler, and T. Schumm, New Journal of Physics **14**, 083019 (2012).

[7] D. Plankensteiner, J. Schachenmayer, H. Ritsch, and C. Genes, Journal of Physics B: Atomic, Molecular and Optical Physics **49**, 245501 (2016).

[8] S. Dubey, G. A. Kazakov, B. Heizenreder, S. Zhou, S. Bennetts, S. A. Schäffer, A. Sitaram, and F. Schreck (MoSaiQC Collaboration), Physical Review Research **7**, 013292 (2025).

[9] R. Kubo, Journal of the Physical Society of Japan **17**, 1100 (1962).

[10] D. Plankensteiner, C. Hotter, and H. Ritsch, Quantum **6**, 617 (2022).


\fi
\end{document}